\documentclass[atoms,article,accept,moreauthors,pdftex]{Definitions/mdpi}

\setitemize{parsep=6pt,itemsep=0pt,leftmargin=*,labelsep=5.5mm}
\setenumerate{parsep=6pt,itemsep=0pt,leftmargin=*,labelsep=5.5mm}
\setlist[description]{itemsep=0mm}

\firstpage{1}
\pubvolume{xx}
\issuenum{1}
\articlenumber{5}
\pubyear{2020}
\copyrightyear{2020}
\history{Received: 31 July 2020; Accepted: 23 September 2020; Published: date}
\updates{yes} 

\usepackage{amssymb}
\usepackage{pdflscape}
\Title{Atomic Data Assessment with PyNeb}

\newcommand{\new}[1]{#1}
\newcommand{\newnew}[1]{{#1}}

\Author{Christophe Morisset $^{1}$\orcidA{},
Valentina Luridiana $^{2,3}$,
Jorge Garc\'ia-Rojas $^{2,3}$\orcidC{}, \\
Ver\'onica G\'omez-Llanos $^{1}$\orcidD{},
Manuel Bautista $^{4}$\orcidE{}, and
Claudio Mendoza $^{4,5,}$*\orcidF{}
}

\AuthorNames{Christophe Morisset, Valentina Luridiana, Jorge Garc\'ia-Rojas, Ver\'onica G\'omez-Llanos, Manuel A. Bautista, Claudio Mendoza, }

\address{%
$^{1}$ \quad Instituto de Astronom\'ia, Universidad Nacional Aut\'onoma de M\'exico, 22860 Ensenada, M\'exico; chris.morisset@gmail.com (C.M.); vgomez@astro.unam.mx (V.G.-L.)\\

$^{2}$ \quad Instituto de Astrof\'isica de Canarias, La Laguna, Tenerife E-38205, Spain; valentina.luridiana@gmail.com (V.L.); jogarcia@iac.es (J.G.-R.)\\

$^{3}$ \quad Departmento de Astrof\'isica, Universidad de La Laguna, La Laguna, Tenerife E-38206, Spain\\

$^{4}$ \quad Department of Physics, Western Michigan University, Kalamazoo, MI 49008, USA; manuel.bautista@wmich.edu\\

$^{5}$ \quad Venezuelan Institute for Scientific Research (IVIC), Caracas 1020, Venezuela
}

\corres{Correspondence: claudio.mendozaguardia@wmich.edu; Tel.: +1-269-387-4935}

\abstract{{\tt PyNeb} is a Python package widely used to model emission lines in gaseous nebulae. \mbox{We take} advantage of its object-oriented architecture, class methods, and historical atomic database to structure a practical environment for atomic data assessment. Our aim is to reduce the uncertainties in the parameter space (line ratio diagnostics, electron density and temperature, and ionic abundances) arising from the underlying atomic data by critically selecting the {\tt PyNeb} default datasets. We evaluate the questioned radiative-rate accuracy of the collisionally excited forbidden lines of the N- and P-like ions (O~{\sc ii}, Ne~{\sc iv}, S~{\sc ii}, Cl~{\sc iii}, and Ar~{\sc iv}), which are used as density diagnostics. With the aid of observed line ratios in the dense NGC~7027 planetary nebula and careful data analysis, we arrive at emissivity ratio uncertainties from the radiative rates within 10\%, a considerable improvement over a previously predicted 50\%. We also examine the accuracy of an extensive dataset of electron-impact effective collision strengths for the carbon isoelectronic sequence recently published. By estimating the impact of the new data on the pivotal [N~{\sc ii}] and [O~{\sc iii}] temperature diagnostics and by benchmarking the collision strength with a measured resonance position, we question their usefulness in nebular modeling. We confirm that the effective-collision-strength scatter of selected datasets for these two ions does not lead to uncertainties in the temperature diagnostics larger than 10\%.}


\begin{document}

\section{Introduction}

{\tt PyNeb}
\footnote{\url{http://research.iac.es/proyecto/PyNeb/}} \cite{lur12a,lur15} is a Python package for the analysis of emission lines in gaseous nebulae, namely H~{\sc ii} regions and planetary nebulae. These lines are used to diagnose the nebular plasma conditions (typical electron densities $n_e\sim 10{-}10^5$~cm$^{-3}$ and temperatures $T_e$ from a few thousand to tens of thousands Kelvin) and to ultimately estimate chemical abundances. Due to reliable observations in the infrared, optical, and ultraviolet and well-constrained plasma models, the ubiquitous galactic and extragalactic nebulae have become abundance tracers of choice to study cosmic chemical evolution \cite{pei17}.

The atomic radiative and collisional rates in nebular models are mainly determined by computation, where inherent theoretical difficulties and computational limitations compromise accuracy. For instance, most collisionally excited lines are dipole forbidden or semi-forbidden ($\Delta S\neq 0$), for which the theoretical $A$-values and collision strengths must take into account fundamental electron-correlation and relativistic effects; additionally, for the forbidden transitions in particular, \mbox{the collision} strengths are dominated by intricate series of resonances \cite{sta12, men14}. For radiative recombination lines, low-temperature dielectronic recombination rates are also sensitive to near-threshold resonances, whose energy positions are model dependent (e.g., $LS$ {vs.} intermediate coupling ionic models) and lack experimental benchmarks \cite{bad15}. It is noteworthy that the development of the codes to calculate relativistic atomic structure and electron--ion scattering has been in great part driven by the stringent requirements of these transition rates, which in many cases are still to converge to the desirable accuracy despite large calculations \cite{pra11}. With the intention of promoting a fluid and constructive interaction between atomic data producers and users in nebular physics, for which we have well-recognized \mbox{contributions \cite{lur11, lur12b},} we have undertaken the present atomic data assessment joint project.

The deep-seated dependence of nebular spectral models on the underlying atomic parameters has encouraged the implementation of databases that have become standard references (e.g., \cite{men83, men20}) or an integral part of spectral modeling tools such as {\sc chianti}\footnote{\url{https://www.chiantidatabase.org/}} \cite{1997Dere_aaps125, der19}, {\sc cloudy}\footnote{\url{https://www.nublado.org/}} \cite{fer17}, {\sc xstar}\footnote{\url{https://heasarc.gsfc.nasa.gov/lheasoft/xstar/xstar.html}} \cite{bau01}, and {\tt PyNeb}. Regarding modeling codes, periodic atomic data curation (i.e., upgrading, selection, and assessment) has become a major determinant of their usefulness, {\tt PyNeb} having the distinctive advantage of keeping a historical database with all the available radiative and collisional datasets rather than discarding data when upgrading. This data-curation strategy allows modelers to estimate the uncertainties of the plasma diagnostics and chemical abundances arising from the scatter of the atomic parameters \mbox{and also} provides a suitable environment for atomic data assessment, the central topic of the present report. This capability of the package is conceptually similar to {\tt AtomPy} \cite{men14b, men20}, but supported by \mbox{a battery} of class methods that facilitate the data revision procedures.

\section{PyNeb Package}

The {\tt PyNeb} Python package \cite{lur15} is tailored to derive the nebular plasma conditions (electron temperature and density) and relative chemical abundances by comparing observed line intensities with computed emissivities, for which it provides a variety of extinction laws, \mbox{ionization correction} factors (ICF), diagnostic plots, and Balmer/Paschen jump temperature determinations. \mbox{Modules, documentation,} historic versions and logs, and notebooks can be accessed from the {\tt PyNeb} {\tt GitHub} repository\footnote{\url{https://github.com/Morisset/PyNeb_devel}}. The current version used in this paper is 1.1.13.

For an $n_{\rm max}$-level ionic model and a given electron temperature and density, {\tt PyNeb} solves the equilibrium equations to determine the level populations, critical densities, and line emissivities, \mbox{which lead} to plasma diagnostics based on observed line ratios. For this purpose, it can upload and manage observational datasets to finally obtain the ionic abundances and, by means of ICF, \mbox{the elemental} abundances. Since the default atomic datasets can be readily changed and updated, \mbox{the package} furnishes plots to compare the atomic parameters from different sources.

Since we are mainly concerned in this report with collisionally excited lines, we recall the basic formalism used by {\tt PyNeb} to compute the emissivities to derive abundances from such lines. Emissivities for collisionally excited lines are obtained by solving the equations of statistical equilibrium for an $n_{\rm max}$-level ionic system:
\begin{equation}
\sum_{k\neq i} n_e n_k q_{ki}+\sum_{k>i} n_k A_{ki} = \sum_{k\neq i} n_e n_i q_{ik}+ \sum_{k<i} n_i A_{ik}\quad\quad\quad
(i=1,n_{\rm max})
\end{equation}
assuming that:
\begin{equation}
\sum_i n_i=n_{\rm tot}\ ,
\end{equation}
where $n_{\rm tot}$ is the total ion density, $n_e$ the electron density, $n_i$ the level population of the $i$th level, \mbox{and $A_{ki}$} the radiative transition rate for levels $k\to i$. The electron-impact de-excitation rate coefficient $q_{ki}$ is usually expressed in terms of the effective collision strength $\Upsilon_{ki}(T_e)$ as:
\begin{equation}
q_{ki}=\frac{8.629\times 10^{-6}}{g_k} \frac{\Upsilon_{ki}(T_e)}{T_e^{1/2}}
\quad\quad (k>i)
\end{equation}
$g_k$ being the statistical weight of the upper level. The excitation rate coefficient can be obtained from the detailed balance relation:
\begin{equation}
q_{ik}=\frac{g_k}{g_i} q_{ki} \exp(-\Delta E_{ik}/k_BT_e)\quad\quad (k>i)
\end{equation}
where $\exp(-\Delta E_{ik}/k_BT_e)$ is the Boltzmann factor. It must be noted that for historic reasons \cite{ost06}, the effective collision strength $\Upsilon_{ki}(T_e)$, i.e., the temperature-dependent Maxwell-averaged collision strength $\Omega_{ki}(E)$, is often referred to in the nebular community simply as the collision strength $\Omega_{ki}(T_e)$.

The line emissivity is given by:
\begin{equation}
\epsilon_{ki}=n_k A_{ki}\, hc/\lambda\ ,
\end{equation}
and for cospatial ions, the emissivity ratio of a pair of optically thin lines is equal to the line intensity ratio, allowing the formulation of plasma diagnostics to determine the electron density and temperature. Furthermore, for an ion $X^a$ emitting a line with wavelength $\lambda$ and intensity $I(\lambda)$, it can be shown that:
\begin{equation}
\frac{n(X^a)}{n({\rm H^+})} = \frac{I(\lambda)}{I({\rm H}\beta)} \frac{\epsilon({\rm H}\beta)}{\epsilon(\lambda)}\ ,
\end{equation}
an expression that leads to the ionic abundance. The elemental abundance is then given by the sum:
\begin{equation}
\frac{n(X)}{n({\rm H})} = \sum_a \frac{n(X^a)}{n({\rm H^+})}=\sum_{a'} \frac{n(X^{a'})}{n({\rm H^+})}\times \hbox{\rm ICF}
\end{equation}
that assumes an ionization correction factor for the unseen ionic species ($a'$ being the set of observed ionic species).

\begin{figure} [H]
 \centering
 \includegraphics[scale=0.4]{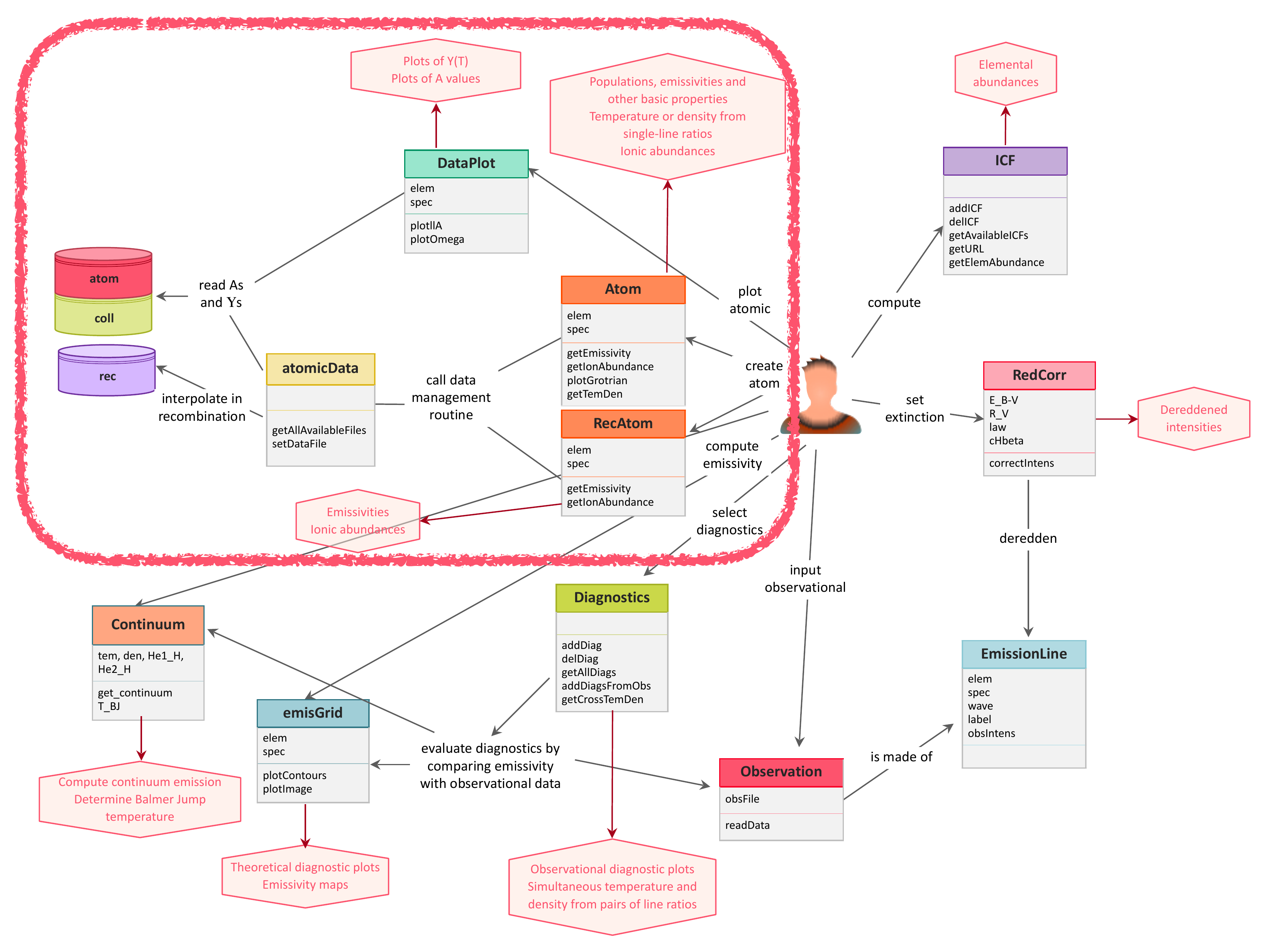}
 \caption{Outline of the {\tt PyNeb} object-oriented architecture showing its classes (name, some arguments, and methods) and output products. For atomic data assessment, we employ the classes and data within the red rectangle. \new{The new {\tt Continuum} class (added in Version 1.1.9) computes the nebular continuum (considering two-photon, free--free, and free--bound emission processes) and the electron temperature associated with the Balmer or Paschen jumps.}}
 \label{pyneb}
 \end{figure}

Figure~\ref{pyneb} gives an outline of the {\tt PyNeb} object-oriented architecture, where the ion object is represented by the {\tt Atom} and {\tt RecAtom} classes. They offer a set of methods to determine the temperature and density ({\tt getTemDen}), emissivities ({\tt getEmissivity}), and ionic abundances ({\tt getionAbundance}) through the {\tt atomicData} manager. In the present work, we use the graphic methods of {\tt DataPlot} to examine the atomic data scatter. The {\tt RecAtom} class computes emissivities for recombination lines by table interpolation or a fitting formula using eponymous methods to obtain the ionic abundance. Useful stacks of $T_e{-}N_e$ emissivity grids can be generated with the {\tt emisGrid} class and multi-diagnostic ratios with {\tt Diagnostics}. Astronomical observations are managed with the {\tt Observation} class, \mbox{and diagnostic} plots can be devised by means of the {\tt Observation}, {\tt Diagnostics}, and {\tt emisGrid} classes.

\section{PyNeb Atomic Database}
\unskip
\subsection{Data File Format}\label{sec:fileformat}

Earlier versions of {\tt PyNeb} relied on data files structured under the FITS format \cite{gre03, pen10} to comply with the IRAF\footnote{\url{https://www.iraf.net/}} {\tt Nebular} package. These FITS files are still included in the current distribution, but are no longer used as default. Since 2014, a new ASCII format was adopted to allow the user easy access to the raw data and the addition of new data if needed. The old FITS files have been transcribed to this ASCII format, and new data are now incorporated in {\tt PyNeb} only abiding this prescription. This format is close to that of the {\tt stout} \cite{fer17} database of the {\sc cloudy} photoionization code, {\tt PyNeb} being able to write its atomic data in {\tt stout} format, but not always able to read from it (especially when different electron temperature grids are used in a single file).
\new{All the files holding the atomic data are downloaded and saved on the user's disk when installing {\tt PyNeb}; they can also be accessed from the {\tt GitHub} repository under the {\tt pyneb/atomic\_data} directory.}

\subsection{Energy Levels}\label{sec:energylevels}

A frequent issue when dealing with complex ions (e.g., Fe$^+$ or Fe$^{++}$) is to be able to compare the atomic parameters corresponding to the same transition, which requires a unique list of energy levels. We have therefore chosen to download the energy levels from the NIST Atomic Spectra Database\footnote{\url{https://www.nist.gov/pml/atomic-spectra-database}} and to sort them in increasing energy order. This may not always correspond to the list of separate multiplets (see, for instance, \cite{1996Zhang_aaps119}) and involves the occasional data reordering to comply with the NIST energy-level order. The user can access the NIST energy levels with the function (for O$^{++}$, say):
\begin{verbatim}
 pn.getLevelsNIST('O3')
\end{verbatim}

\subsection{Recombination Spectra for H and He}\label{sec:recombination}

Recombination emissivities are read from the {\it{x}}\_{\tt rec}\_{\it{ref}}.{\it ext} files, where $x$ represents the ion and {\it ref} the source reference acronym. The {\it ext} extension may be either {\tt func} or {\tt hdf5}; for the former, a numerical function is used as in \cite{1991Pequignot_aap251}. The data are read by the {\tt RecAtom} class and interpolated in density and temperature. Recently, recombination data for the heavier ions of C, O, N, and Ne have also been incorporated into the {\tt PyNeb} atomic database. For O$^{++}$, the recombination data table from \cite{2017Storey_mnra470} is split into infrared and optical--ultraviolet lines, and the A, B, and C cases are stored in separate files.

\subsection{Collisional Spectra}\label{sec:collisional}

For collisional spectra, {\tt PyNeb} creates an $n_\mathrm{max}$-level atom model characterized by the atomic data read at runtime from the flat ASCII files $xxx$\_{\tt atom}\_{\it{ref}}.{\tt dat} and $xxx$\_{\tt coll}\_{\it{ref}}.{\tt dat}, where $xxx$ represents the ion (e.g., ``o\_iii'' for O$^{++}$) and {\it ref} the source reference acronym. Transition probabilities $A_{ki}$ (s$^{-1}$) are different from zero only for $k > i$, so the data are ordered in a rectangular $n_\mathrm{max}\times n_\mathrm{max}$ matrix with zeroes along and above the main diagonal. The {\tt atom} files contain the following items:

\begin{itemize}
\item{}Transition probabilities.
\item{}Information about the corresponding ion.
\item{}The data source reference.
\end{itemize}

The {\it{x}}\_{\tt coll}\_{\it{ref}}.{\tt dat} file contains the effective collision strengths (ECS) $\Upsilon_{ki}(T_e)$, which are functions of the electron temperature (i.e., a collective property of the electron distribution) obtained by averaging the energy-dependent collision strengths $\Omega_{ki}(E)$ over a Maxwellian distribution of electron \mbox{kinetic energies.}

The ECS are usually published in tabular form for a handful of temperatures $T_e$ and must be interpolated to get the $\Upsilon_{ki}(T_e)$ at the desired $T_e$ value. Such a table is included in the data file, \mbox{and since} for each transition $k\rightarrow i$, the upper level $k > i$, there is a whole one-dimensional vector array of collision strengths with one element for each tabulated temperature value; that is, the complete $\Upsilon$ set for an atom is a three-dimensional array that prevents confusing the data in the matrix with the transition probabilities (each transition is presented instead in a line). The temperature can be in Kelvin, log10(K), or K/10,000 depending on the particular dataset. The $\Upsilon_{ki}(T_e)$ are interpolated linearly between the values at two contiguous temperatures; a keyword can be used to invoke an alternative interpolation method depending on what is available from the {\tt interpolate.interp1d} function of the {\tt scipy} library (e.g., ``quadratic'' or ``cubic''). The interpolation method has some second-order influence on the resulting ECS, and thus on the emissivities.

The {\tt coll} files contain the following quantities:

\begin{itemize}
\item{} Electron-temperature values used for the grid.
\item{} ECS at the corresponding temperatures.
\item{} Information about the corresponding ion (and the unit used for the electron temperature).
\item{} Data source reference.
\end{itemize}

\section{Data from CHIANTI}

{\tt PyNeb} can read the atomic data from the {\sc chianti}\footnote{\new{\url{https://www.chiantidatabase.org/}}} package (Versions 7, 8, and 9, the latter the latest release; see \cite{1997Dere_aaps125, der19}). The user needs the {\sc chianti} database installed beforehand on their computer and the {\tt XUVTOP} environment variable pointing to the corresponding directory. The {\sc chianti} data (namely, level energies $E$, transition probabilities $A$, and ECS $\Upsilon(T_e)$) can be accessed \new{using commands based on {\tt Chiantipy} routines included in the {\tt PyNeb} package} (for O$^{++}$ at 10,000~K, say):
\begin{verbatim}
 pn.utils.pn_chianti.Chianti_getA('o_3')
 pn.utils.pn_chianti.Chianti_getOmega('o_3', tem = 1e4)
\end{verbatim}
The correspondence
 between the NIST and {\sc chianti} level orders is obtained with the command:
\begin{verbatim}
 pn.utils.pn_chianti.get_levs_order('O3', NLevels = 9)
 -> {0: 0, 1: 1, 2: 2, 3: 3, 4: 4, 5: 5, 6: 8, 7: 6, 8: 7}
\end{verbatim}
where the first number is the NIST index and the second the corresponding {\sc chianti} index. We can see that Levels 7, 8, and 9 are not in the same order in both databases.

In {\sc chianti}, the effective collision strength is mapped onto a finite temperature interval $(0,1)$ with the scaling relations of \cite{bur92} and represented with a spline of five \new{or more points. This ensures that \mbox{a value} can be obtained at any temperature $(0,\infty)$. From Version 8, the mapping onto the $(0,1)$ \newnew{scaled} temperature interval is still used,} \newnew{but the stored points now correspond to the original data replacing the spline fits by interpolation}. \new{However, this only applies to new data; for the data already in the database, the 5-, 7-, or 9-point spline fits are still used \cite{2015Del-Zanna_aap582}; e.g., a 5-point spline for O$^{++}$.} \newnew{Although this concise representation may not always be accurate, e.g., at low temperatures \cite{2015Del-Zanna_aap582}}, it has endured and has been extended to neutral atoms and molecules \cite{sum00}. \new{In {\tt PyNeb}, the original data are used. The extrapolation to low and high temperatures is performed by reporting the lowest and highest value, respectively. The user can avoid extrapolation by adding {\tt noExtrapolation = True} when instantiating the {\tt Atom} object; in this case, {\tt NaN} values are returned for temperatures outside of the original range.}

\section{Atomic Data File Management}
\new{Within the scope of {\tt PyNeb} in atomic data assessment, we go over the perhaps unfamiliar but core concept of the atomic data dictionary and its default value, the latter being the result of data selection procedures we are hereby trying to refine. We explain how to list the default and other dictionaries, the dataset labels, and respective bibliographic references and how to construct bespoke dictionaries to devise data comparisons.}

\subsection{Atomic Data Files}

The collection of data files for the different ions used in \mbox{a calculation} is organized in \mbox{a dictionary} (in the Python sense); i.e., a data structure with one or more entries such that each entry has \mbox{a uniquely} defined key and value. An atomic data dictionary is identified by a label, \mbox{which can} be either predefined (i.e., built-in) or provided by the user; the keys are the ions, and the values specify the atomic data. More precisely, each value is itself a dictionary with three possible keys, \mbox{``atom'', ``coll'', or ``rec'', }\mbox{whose respective} values are the names of the atomic, \mbox{collisional, or recombination} data files.

The default dictionary of the current atomic database is obtained with the command:
\begin{verbatim}
 pn.atomicData.defaultDict
\end{verbatim}
At present, it matches {\tt 'PYNEB\_20\_01'}, but it will change as new atomic data are published or our criteria for selecting datasets evolves. The list of all the built-in atomic data labels is obtained with:
\begin{verbatim}
 pn.atomicData.getAllPredefinedDict()
\end{verbatim}
At the time of writing, 11 such dictionaries exist: {\tt 'IRAF\_09\_orig'}, {\tt 'IRAF\_09'}, {\tt 'PYNEB\_13\_01'}, {\tt 'PYNEB\_14\_01'}, {\tt 'PYNEB\_14\_02'}, {\tt 'PYNEB\_14\_03'}, {\tt 'PYNEB\_16\_01'}, {\tt 'PYNEB\_17\_01'}, {\tt 'PYNEB\_17\_02'}, {\tt 'PYNEB\_18\_01'}, and {\tt 'PYNEB\_20\_01'}.
For the dictionary {\tt 'PYNEB\_14\_01'} and earlier, old FITS files are used, and a special command must be run before using them: {\tt pn.atomicData.includeFitsPath()}.
The old dictionaries are kept in {\tt PyNeb} to guarantee backward compatibility, but users are strongly urged to use the default and to quote the {\tt PyNeb} version in publication. As an example, the entry corresponding to the current O$^{++}$ data is:
\begin{verbatim}
 pn.atomicData.getDataFile(atom = 'O3')

 'o_iii_atom_FFT04-SZ00.dat',
 'o_iii_coll_SSB14.dat',
 'o_iii_rec_P91.func'
\end{verbatim}
If no value (or {\tt None}) is specified for the atom keyword, the list of the data files for all the ions of the current configuration is returned as a dictionary.

The bibliographic references of the atomic data in the default dictionary are displayed with:
\begin{verbatim}
 pn.atomicData.printAllSources()
\end{verbatim}
The bibliographic references of the atomic data used in a given dictionary labeled {\tt ``label''} (default or other) can be displayed by entering:
\begin{verbatim}
 pn.atomicData.printAllSources(predef = 'PYNEB_16_01')
\end{verbatim}
The bibliographic references of the atomic data used for just a subset of ions (e.g., O{\sc~ii}, O{\sc~iii}, and S{\sc~ii}) can be shown with:
\begin{verbatim}
 pn.atomicData.printAllSources(at_set = ['O2', 'O3', 'S2'])
\end{verbatim}

\subsection{Changing an Individual Data File or the Whole Dataset}

The data available for a given ion (e.g., O{\sc~iii}) can be displayed by entering:

\begin{verbatim}
 pn.atomicData.getAllAvailableFiles('O3')

 ['* o_iii_atom_FFT04-SZ00.dat',
 '* o_iii_coll_SSB14.dat',
 '* o_iii_rec_P91.func',
 'o_iii_atom.chianti',
 'o_iii_atom_FFT04.dat',
 'o_iii_atom_GMZ97-WFD96.dat',
 'o_iii_atom_SZ00-WFD96.dat',
 'o_iii_atom_TFF01.dat',
 'o_iii_atom_TZ17.dat',
 'o_iii_coll.chianti',
 'o_iii_coll_AK99.dat',
 'o_iii_coll_LB94.dat',
 'o_iii_coll_MBZ20.dat',
 'o_iii_coll_Pal12-AK99.dat',
 'o_iii_coll_TZ17.dat']
\end{verbatim}
The data used to currently instantiate the {\tt Atom} and {\tt RecAtom} objects are identified with a leading asterisk (*). The current file selection for a given ion can be modified, for example, with the commands:
\begin{verbatim}
 pn.atomicData.setDataFile('o_iii_atom_TZ17.dat')
 pn.atomicData.setDataFile('o_iii_coll_TZ17.dat')
\end{verbatim}
The next instantiating of an {\tt Atom} object for O$^{++}$ will use these data files. It is then simple to instantiate different {\tt Atom} objects for the same ion with different atomic data to compare, for example, emissivities.

\subsection{Deprecated Data}

It may happen that at some point in the history of {\tt PyNeb}, some atomic data files are removed from the default directory where all the data are stored. This mainly occurs if the same data are defined in another file that takes better into account the original source of the atomic data; in this case, the compilation is not used anymore (e.g., the data that used to be in the {\tt s\_iii\_atom\_PKW09.dat} file is now in the {\tt s\_iii\_atom\_FFTI06.dat} file to cite the original paper). When this occurs, the obsolete file is sent to the {\tt deprecated} directory to ensure backward compatibility. If the user still prefers to have access to the data in this directory, the command {\tt pn.atomicData.includeDeprecatedPath()} is needed before permuting the default data file with the deprecated one. Removing deprecated files from the accessible path is performed with {\tt pn.atomicData.removeDeprecatedPath()}.

\subsection{List of All the Data Used in a Script}

Each time an {\tt Atom} or a {\tt RecAtom} object is instantiated, the data files used are stored in a dictionary held by the {\tt atomicData} object. After a script is run using {\tt PyNeb}, the list of all the files can be obtained using the {\tt pn.atomicData.usedFiles} command. \new{This helps the user cite the atomic data \mbox{papers properly.}}

\section{Continuum Class: Balmer and Paschen Jumps to Determine the Electron Temperature}

\new{{\tt Continuum} is a new class in {\tt PyNeb} (see Figure~\ref{pyneb}) devised to obtain alternative estimates of the electron temperature. This is an important development as the temperatures derived from line diagnostics strongly depend on the basic atomic database, whose accuracy is often questioned in the literature (e.g., \cite{nic13}).}

\subsection{Nebular Continuum}\label{nebcon}
\begin{figure}[H]
 \centering
 \includegraphics[scale = 0.6]{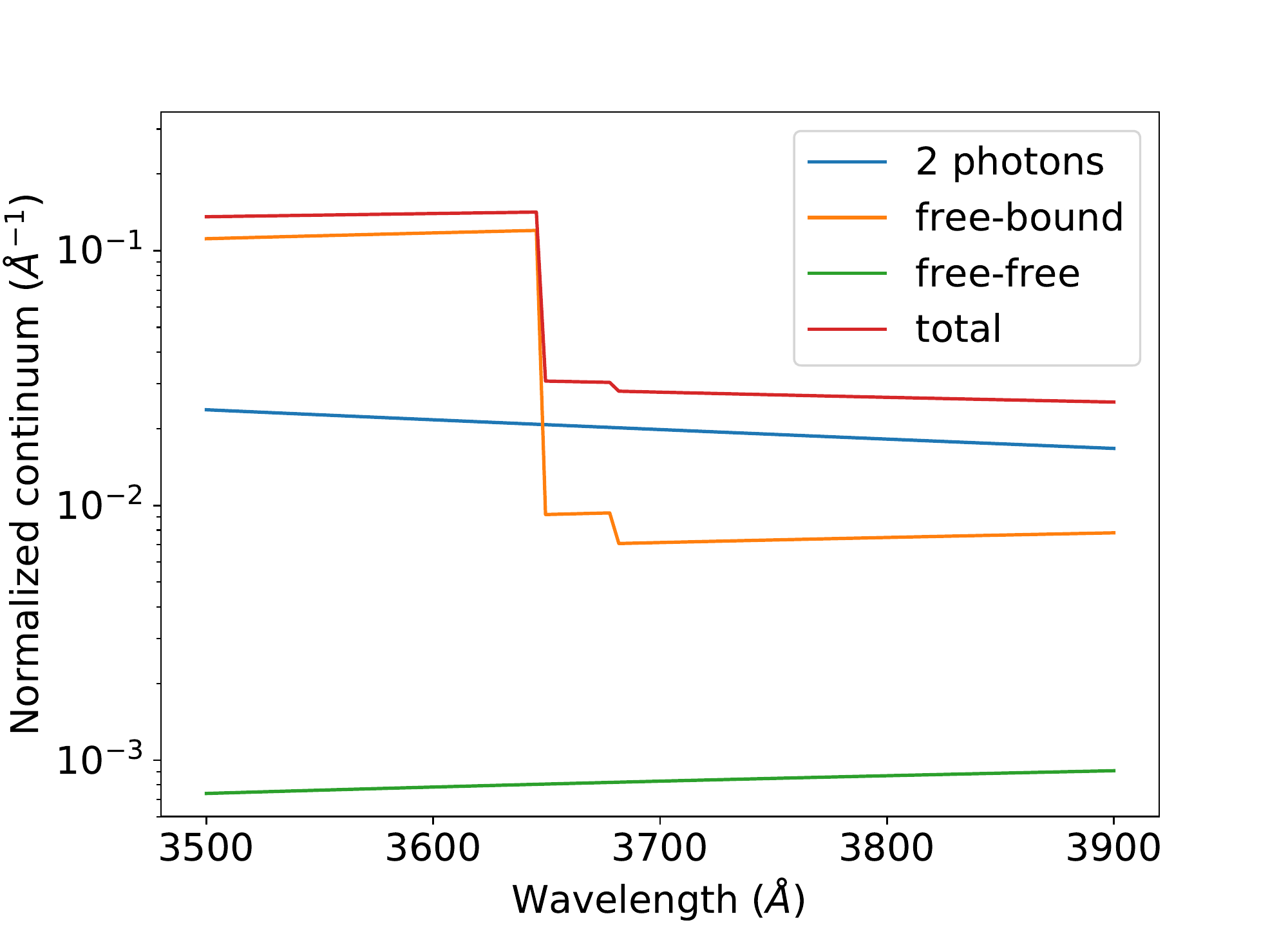}
 \caption{Nebular continuum normalized to H~{\sc i}~3770\,\AA. The data are obtained with the {\tt get\_continuum} function of the {\tt Continuum} class for $T_e=10^4$~K, $n_e=10^2$~cm$^{-3}$, He$^+$/H$^+ =0.12$, and He$^{++}$/H$^+=0.01$.} \label{fig:continuum}
\end{figure}

The nebular continuum due to the following processes,
\begin{itemize}
 \item Free--bound emission from H$^{+}$, He$^{+}$, and He$^{++}$ \cite{2006Ercolano_mnras372},
 \item Free--free emission by H$^{+}$, He$^{+}$, and He$^{++}$ \cite{1991Storey_CoPhC66},
 \item Two-photon decay from 2$^2$S of H \cite{ost06},
\end{itemize}
can be estimated with {\tt PyNeb}. The prescription to obtain the continuum is: wavelength range; \mbox{electron temperature} and density; and the He$^+$/H$^+$ and He$^{++}$/H$^+$ ionic abundances. The continuum is given in erg\,s$^{-1}$\,cm$^{-3}$\AA$^{-1}$ or in \AA$^{-1}$ when normalized to an H{\sc~i} line. The commands:
\begin{verbatim}
 C = pn.core.continuum.Continuum()
 C.get_continuum(tem = 1e4, den = 1e2, He1_H = 0.12, He2_H = 0.01,
    wl = np.linspace(3500, 3900, 100), HI_label = '11_2')
\end{verbatim}
give an example on how to compute the continuum normalized to H~{\sc i}~3770\,\AA\ for: $T_e=10^4$~K; \mbox{$n_e=10^2$~cm$^{-3}$;} $\mathrm{He}^+/\mathrm{H}^+ = 0.12$; and $\mathrm{He}^{++}/\mathrm{H}^+=0.01$ in the wavelength range 3500--3900~\AA. \mbox{The result} of this example is plotted in Figure~\ref{fig:continuum}. To get the continuum with no normalization, \mbox{{\tt HI\_label}} must be defined as {\tt None}.

\subsection{Balmer and Paschen Jump Temperatures}

The electronic temperature from the Balmer (or Paschen) jump is determined by minimizing the difference between the theoretical nebular continuum (described in Section~\ref{nebcon}) and the observed flux before and after the jump. The optimization is based on the Richard Brent algorithm for \mbox{root-finding \cite{1973Brent}} using {\tt scipy.optimize.brentq}.

The temperature determination requires: the difference of the observed flux (in \AA$^{-1}$) before and after the jump normalized to a H~{\sc i} line; the corresponding wavelengths (predefined at 3643 and 3861~\AA); the electron density; and the He$^+$/H$^+$ and He$^{++}$/H$^+$ ionic abundances. The following example uses the data for the planetary nebula M~1-42 \cite{2001Liu_mnras327} normalized to H~{\sc i}~3770\,\AA\ and to H$\beta$:
\begin{verbatim}
 C.T_BJ(BJ_HI = 0.23, den = 1500, He1_H = 0.139, He2_H = 0.009,wl_bbj = 3643,
  wl_abj = 3861, HI_label = '11_2', T_min = 500.0, T_max = 30000.0)

 3748.803973930705
\end{verbatim}

\begin{verbatim}
 C.T_BJ(BJ_HI = 8.56e-3, den = 1500, He1_H = 0.139, He2_H = 0.009,
  wl_bbj = 3643, wl_abj = 3861, HI_label='4_2', T_min = 500.0, T_max = 30,000.0)

 3910.6448636964446
\end{verbatim}
The contributions from the stellar continuum and dust emission are not considered in the temperature determination, although they can contribute to the observed flux.

\section{Atomic Data Assessment}

A paradigmatic example of using {\tt PyNeb} for atomic data assessment is \cite{jua17}, who analyzed the size of the systematic uncertainties when using 52 different sets of transition probabilities and collision strengths to calculate the physical conditions and total abundances of O, N, S, Ne, Cl, and Ar for \mbox{a sample} of planetary nebulae and H~{\sc ii} regions. These authors found that at low densities, the scatter in the abundance ratios was lower than 0.1--0.2~dex\footnote{Contraction of decimal exponent: 1~dex means an order of magnitude or a factor of 10.}, but for densities $n_e > 10^4$~cm$^{-3}$, it can be larger than 0.6--0.8~dex for several abundance ratios such as O/H and N/O. A narrower scatter was obtained by discarding questionable atomic data, but a revision of the atomic data behind important density diagnostics was recommended: the radiative transition probabilities ($A$-values) of O~{\sc ii}, S~{\sc ii}, Cl~{\sc iii}, \mbox{and Ar~{\sc iv}}, as well as the effective collision strengths for the latter.

In the last few years, there has been growing interest in determining chemical abundances in ionized nebulae for the heavier elements nucleosynthesized by slow-neutron capture (the s-process). Encouraged by several detections of s-process faint emission lines in planetary nebulae, several groups have been computing the required atomic data; e.g., Rb~{\sc iv} \cite{ster16}; Se~{\sc iii} and Kr~{\sc vi} \cite{ster17}; \mbox{and Te~{\sc iii}} \mbox{and Br~{\sc v} \cite{mad18}.} The synergy between observations, atomic data calculations, and numerical modeling (needed to determine ICF) has significantly boosted this field. All these atomic datasets \new{for neutron-capture elements (energy levels, transition probabilities, and collision strengths)} have been incorporated in {\tt PyNeb}, allowing, for the first time, total abundance estimates for Kr and Se by using the complete sets of ICF developed for these elements by \cite{ster17, mad18} \new{using a private ad hoc version \mbox{of {\sc cloudy}}.}

\begin{figure}[H]
 \centering
 \includegraphics[width=0.9\textwidth]{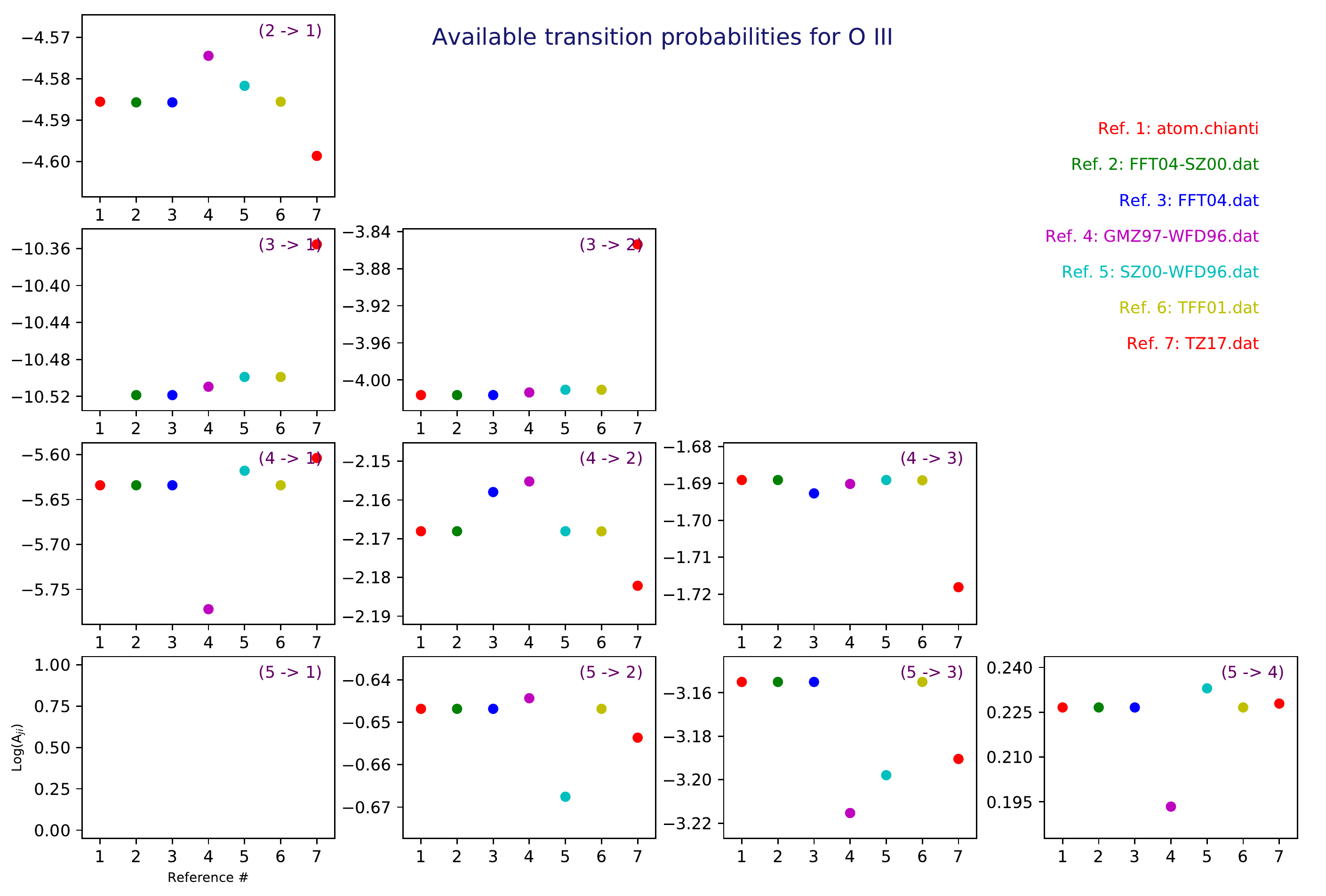}
 \caption{O~{\sc iii} $A$-value diagram obtained with the {\tt plotAllA} method of the {\tt DataPlot} class.}
 \label{Avalues}
\end{figure}

The {\tt DataPlot} class has several methods to plot the atomic datasets, which can be used to bring out the data scatter and questionable outliers. For instance, the commands:
\begin{verbatim}
 dp_O3=pn.DataPlot('O',3,NLevels = 5)
 dp_O3.plotAllA(figsize = (10,8))
\end{verbatim}
plot all the available $A$-values for O~{\sc iii} as shown in Figure~\ref{Avalues}.

To further illustrate the possibilities of the {\tt PyNeb} analysis tools in atomic data assessment, \mbox{we discuss} in the next sections two relevant study cases.

\subsection{A-Values for N- and P-Like Ions}

\begin{figure}[H]
 \centering
 \includegraphics[width=0.7\textwidth]{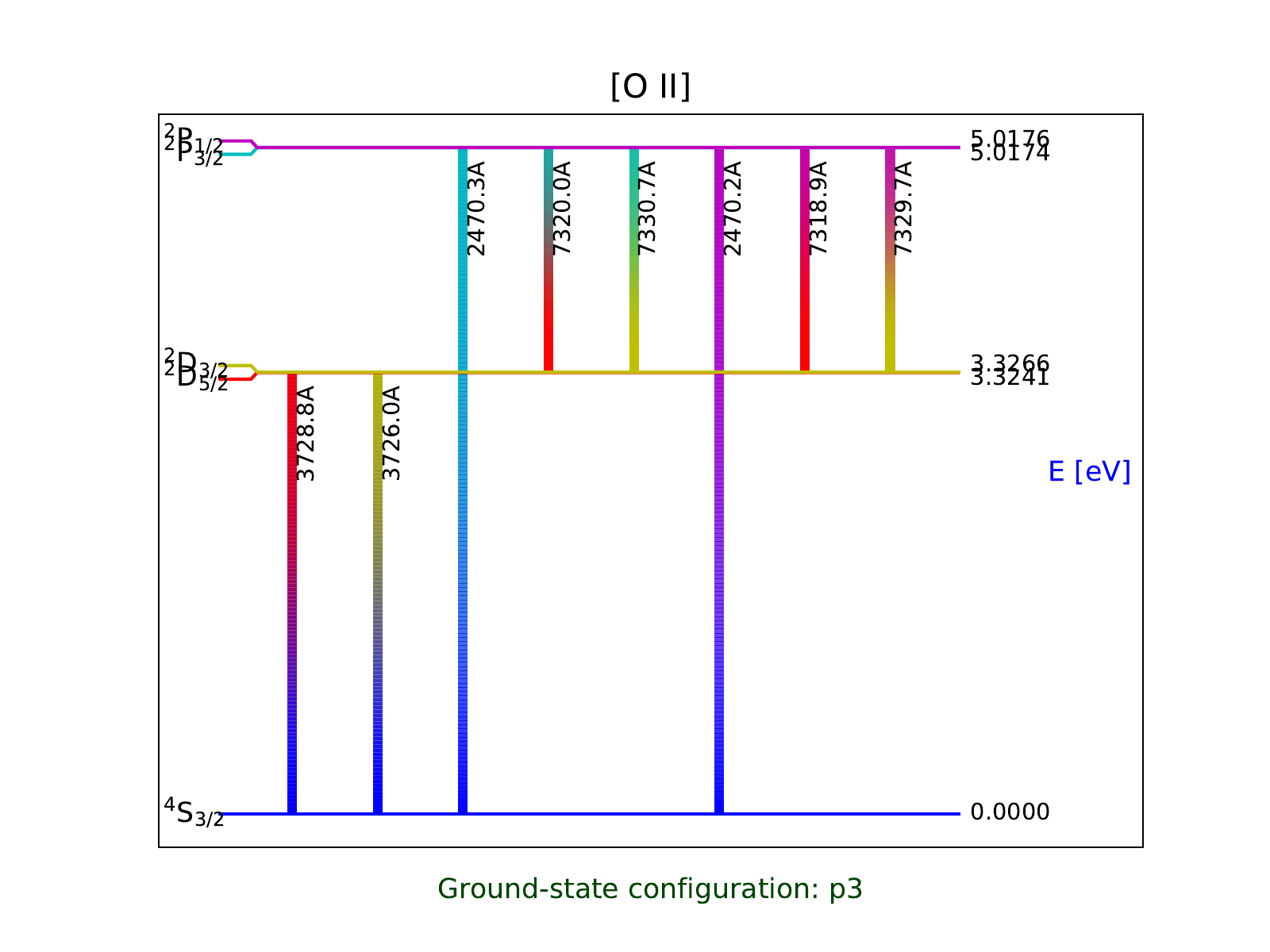}
 \caption{O~{\sc ii} Grotrian diagram obtained with the {\tt plotGrotrian} method of the {\tt Atom} class.}
 \label{fig:o2_levels}
\end{figure}

We address here one of the atomic data queries raised by \cite{jua17}: the radiative rates for the N-like (O~{\sc ii} and Ne~{\sc iv}) and P-like (S~{\sc ii}, Cl~{\sc iii}, and Ar~{\sc iv}) systems, which are widely used to devise nebular density diagnostics. These are ionic species with $n\mathrm{s}^2n\mathrm{p}^3$ ground configuration ($n=2$ for N-like and $n=3$ for P-like), the line ratio of interest being the $^2\mathrm{D^o_{3/2,5/2}}\rightarrow\, ^4\mathrm{S^o_{3/2}}$ doublet.
Figure~\ref{fig:o2_levels} shows the energy-level structure and radiative transitions of O~{\sc ii} plotted with the {\tt PyNeb} commands:
\begin{verbatim}
 O2 = pn.Atom('O',2)
 O2.plotGrotrian()
\end{verbatim}
in particular the 3729/3726 density-sensitive doublet.

In the context of nebular physics, the accuracy of the $A$-values for O~{\sc ii} and S~{\sc ii} have been previously discussed \cite{sta12, 2013Stasinska_aap552, men14}. For data assessment purposes, the $A$-value ratio:
\begin{equation}
R_1=\frac{3}{2}\times\frac{A(^2\mathrm{D^o_{5/2}}-\,^4\mathrm{S^o_{3/2}})}{A(^2\mathrm{D^o_{3/2}}-\,^4\mathrm{S^o_{3/2}})}
\end{equation}
was therein prescribed to provide a useful observational benchmark since:
\begin{equation}
 \frac{I(^2\mathrm{D^o_{5/2}}-\,^4\mathrm{S^o_{3/2}})}{I(^2\mathrm{D^o_{3/2}}-\,^4\mathrm{S^o_{3/2}})}\longrightarrow R_1 \quad\hbox{\rm as}\quad N_e\rightarrow\infty\ ;
\end{equation}
that is, at high densities, the line intensity ratio depends mostly on the doublet $A$-values, and therefore physically:
\begin{equation}\label{rat}
\frac{I(^2\mathrm{D^o_{5/2}}-\,^4\mathrm{S^o_{3/2}})}{I(^2\mathrm{D^o_{3/2}}-\,^4\mathrm{S^o_{3/2}})} \geq R_1\ .
\end{equation}

\begin{figure}[H]
 \centering
 \includegraphics[width=0.4\textwidth]{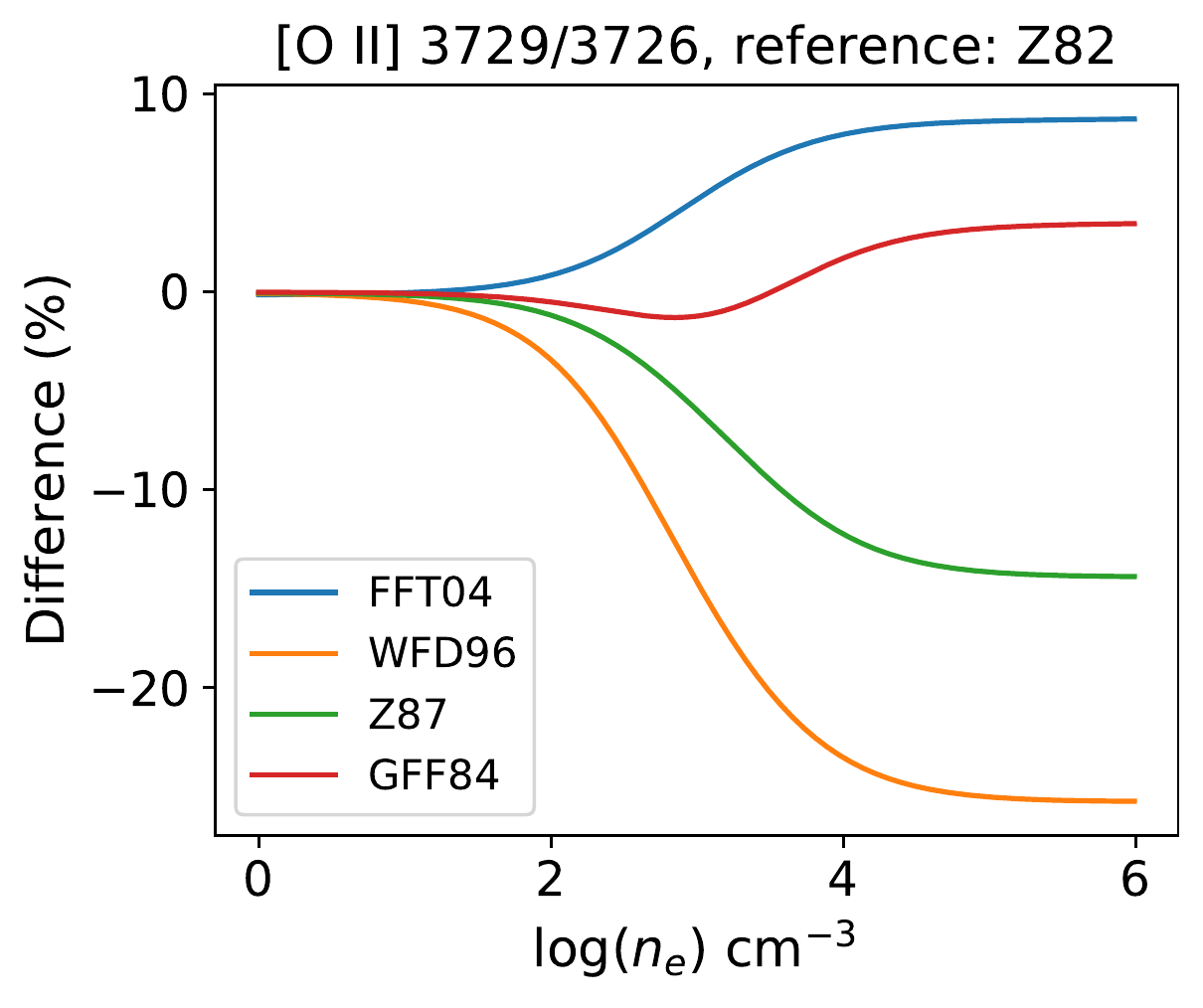}
 \includegraphics[width=0.4\textwidth]{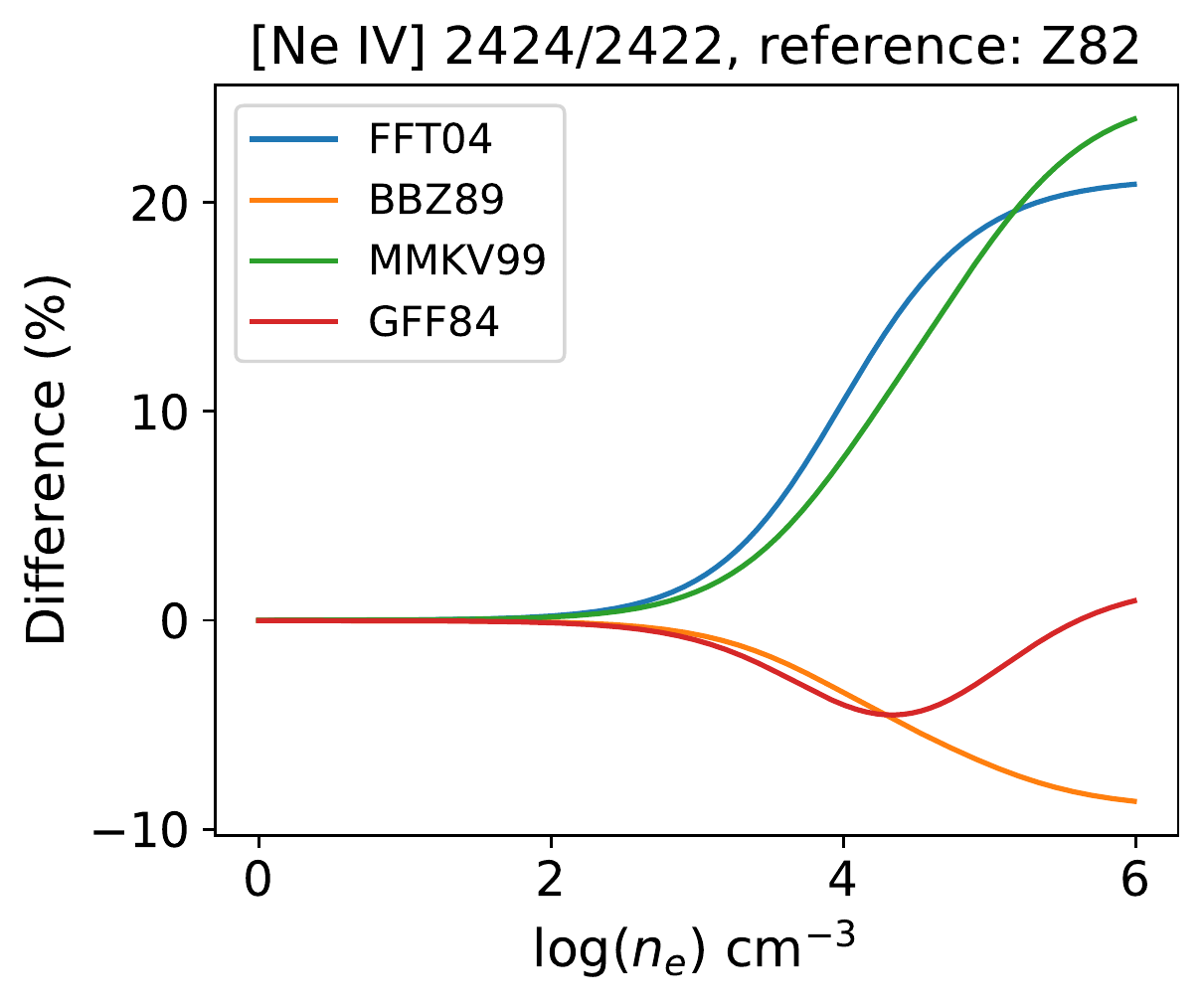}
 \caption{Comparison of $\varepsilon(^2\mathrm{D^o_{5/2}}-\,^4\mathrm{S^o_{3/2}})/\varepsilon(^2\mathrm{D^o_{3/2}}-\,^4\mathrm{S^o_{3/2}})$ density-sensitive emissivity ratio of N-like ions computed at $T_e=10^4$~K with different $A$-value datasets relative to Z82. {Left panel}: [O~{\sc ii}] $\varepsilon(\lambda 3729)/\varepsilon(\lambda 3726)$. {Right panel}: [Ne~{\sc iv}] $\varepsilon(\lambda 2424)/\varepsilon(\lambda 2422)$. Source references for the $A$-value datasets are given in Table~\ref{nlike}.\label{nlike_dif}}
\end{figure}

Using the {\tt getEmissivity} method, we plot in Figure~\ref{nlike_dif} the percentage difference of the \mbox{$\varepsilon(^2\mathrm{D^o_{5/2}}-\,^4\mathrm{S^o_{3/2}})/\varepsilon(^2\mathrm{D^o_{3/2}}-\,^4\mathrm{S^o_{3/2}})$} emissivity ratio in N-like ions at $T_e=10^4$~K computed with $A$-values from the GFF84 \cite{god84}, Z87 \cite{zei87}, WFD96 \cite{wie96}, and FFT04 \cite{fro04} datasets relative to Z82 \cite{zei82}. All these calculations have been performed with a relativistic Breit--Pauli Hamiltonian; they account for electron correlation effects, and some include corrections to the magnetic dipole operator and level energy separations to increase accuracy. Significant discrepancies due to the $A$-value choice begin to appear at electron densities $\log(n_e)\gtrsim 2$~cm$^{-3}$, and most grow with density until $R_1$ is reached. \mbox{For both} O~{\sc ii} and Ne~{\sc iv}, the emissivity-ratio uncertainties from the radiative data are no larger than 25\% along the entire density range.

A further measure of $A$-value accuracy can be carried out in terms of the $R_1$ $A$-value ratio and a benchmark with the observed line ratios from the high-density NGC~7027 planetary \mbox{nebula \cite{zha05}} as listed in Table~\ref{nlike}. The overall agreement between the theoretical $R_1$ ratios is around the 15\% level. Interestingly, the more recent calculations (CQL07 \cite{che07}, HGZJYL14 \cite{han14}, and \mbox{HLZSZ18 \cite{he18}}) have focused $A(^2\mathrm{D^o_{J}}-\,^4\mathrm{S^o_{3/2}})$ to constrain $R_1$ using a fully relativistic multi-configuration Dirac--Fock (MCDF) method with convergent configuration-interaction expansions, which includes the Breit interaction and QED effects. Although the $R_1$ from these calculations is within the general scatter, \mbox{the absolute} $A(^2\mathrm{D^o_{5/2}}-\,^4\mathrm{S^o_{3/2}})$ values from these calculations are somewhat larger than \mbox{previous estimates.}

\begin{table}[H]
 \caption{Theoretical $A$-values (s$^{-1}$) for the $^2\mathrm{D^o_{3/2,5/2}}-\,^4\mathrm{S^o_{3/2}}$ doublet within the $\mathrm{2s^22p^3}$ ground configuration of N-like O~{\sc ii} and Ne~{\sc iv}. The $R_1$ ratio is also given for comparison with the line intensity ratio observed in the dense NGC~7027 planetary nebula. \label{nlike}}
 \centering
 \begin{tabular}{cccccc}
 \toprule
 \textbf{Ion} & \textbf{Dataset} & \boldmath{$A(^2\mathrm{D^o_{3/2}}-\,^4\mathrm{S^o_{3/2}})$} & \boldmath{$A(^2\mathrm{D^o_{5/2}}-\,^4\mathrm{S^o_{3/2}})$} & \boldmath{$R_1$} & \textbf{Obs Ratio}\\
 \midrule
 O~{\sc ii} & Z82 \cite{zei82} & 1.65 $\times$ 10$^{-4}$ & 3.82 $\times$ 10$^{-5}$ & 3.48 $\times$ 10$^{-1}$ &  \\
   & GFF84 \cite{god84} & 1.50 $\times$ 10 $^{-4}$ & 3.59 $\times$ 10$^{-5}$ & 3.59 $\times$ 10$^{-1}$ &  \\
   & Z87 \cite{zei87} & 1.81 $\times$ 10$^{-4}$ & 3.59 $\times$ 10$^{-5}$ & 2.97 $\times$ 10$^{-1}$ &  \\
   & WFD96 \cite{wie96} & 1.78 $\times$ 10$^{-4}$ & 3.06 $\times$ 10$^{-5}$ & 2.58 $\times$ 10$^{-1}$ &  \\
   & FFT04 \cite{fro04} & 1.64 $\times$ 10$^{-4}$ & 4.12 $\times$ 10$^{-5}$ & 3.78 $\times$ 10$^{-1}$ &  \\
   & CQL07
 \cite{che07} & $1.83^{+0.03}_{-0.10}$ $\times$ 10$^{-4}$ & $4.21^{+0.09}_{-0.11}$ $\times$ 10$^{-5}$ & $3.45^{+0.28}_{-0.14}$ $\times$ 10$^{-1}$ & \\
   & HGZJYL14
 \cite{han14} & 1.76 $\times$ 10$^{-4}$ & 4.24 $\times$ 10$^{-5}$ & 3.62 $\times$ 10$^{-1}$ &  \\
   & HLZSZ18
 \cite{he18} & 1.75 $\times$ 10$^{-4}$ & 4.34 $\times$ 10$^{-5}$ & 3.72 $\times$ 10$^{-1}$ &  \\
   & NGC 7027 \cite{zha05} &  &  &  & 3.61 $\times$ 10$^{-1}$ \\
 Ne~{\sc iv} & Z82 \cite{zei82} & 5.54 $\times$ 10$^{-3}$ & 4.84 $\times$ 10$^{-4}$ & 1.31 $\times$ 10$^{-1}$ &  \\
   & GFF84 \cite{god84} & 4.97 $\times$ 10$^{-3}$ & 4.41 $\times$ 10$^{-4}$ & 1.33 $\times$ 10$^{-1}$ &  \\
   & BBZ89 \cite{bec89} & 5.77 $\times$ 10$^{-3}$ & 4.58 $\times$ 10$^{-4}$ & 1.19 $\times$ 10$^{-1}$ &  \\
   & MMKV99 \cite{mer99} & 4.97 $\times$ 10$^{-3}$ & 5.43 $\times$ 10$^{-4}$ & 1.64 $\times$ 10$^{-1}$ &  \\
   & FFT04 \cite{fro04} & 5.50 $\times$ 10$^{-3}$ & 5.82 $\times$ 10$^{-4}$ & 1.59 $\times$ 10$^{-1}$ &  \\
   & HGZJYL14 \cite{han14} & 5.56 $\times$ 10$^{-3}$ & 5.02 $\times$ 10$^{-4}$ & 1.36 $\times$ 10$^{-1}$ &  \\
   & HLZSZ18 \cite{he18} & 5.59 $\times$ 10$^{-3}$ & 5.18 $\times$ 10$^{-4}$ & 1.39 $\times$ 10$^{-1}$ &  \\
   & NGC 7027 \cite{zha05} &  &  &  & 3.42 $\times$ 10$^{-1}$ \\
 \bottomrule
 \end{tabular}
\end{table}

It may be seen in Table~\ref{nlike} that the observed [O~{\sc ii}] line ratio is in good agreement with the theoretical $R_1$ values, while in Ne~{\sc iv}, it is noticeably higher, implying that the high-density regime for this latter diagnostic has not been reached in NGC~7027, and it can therefore be used to obtain the electron density. We deduce an electron density of $\log(n_e)=4.7\pm 0.1$~cm$^{-3}$ at $T_e=10^4$~K using the Z82, \mbox{GFF84, BBZ89, MMKV99,} and FFT04 datasets, giving an estimate of the impact of the atomic data scatter on the density diagnostic, namely 0.1~dex. Since one of the main objectives of the present work is to select default atomic datasets in {\tt PyNeb}, we singled out Z82 and GFF84 for both O~{\sc ii} and Ne~{\sc iv} after much pondering to ensure line ratio diagnostics within 10\%. For this purpose, we applied the following selection criteria:

\begin{itemize}
 \item Datasets must contain all the transitions within the ground configuration with at least $\Delta L\neq 0$.
 \item For consistency, datasets considering isoelectronic sequences are preferred to those focusing on \mbox{single species.}
 \item Wavelength adjusted $A$-values computed with correct transition operators (e.g., magnetic dipole) \mbox{have priority.}
 \item Emissivity ratios must lie within a 10\% scatter along the isoelectronic sequence.
 \item The dataset must comply with the condition of Equation~(\ref{rat}), and $R_1$ must lie within a 10\% \mbox{theoretical scatter.}
 \item A selected dataset must be validated with data computed independently with a different \mbox{numerical method.}
\end{itemize}
Among the Z82, GFF84, BBZ89, MMKV99, and FFT04 datasets, only the first two comply with these selection criteria.

\begin{figure}[H]
 \centering
 \includegraphics[width=0.3\textwidth]{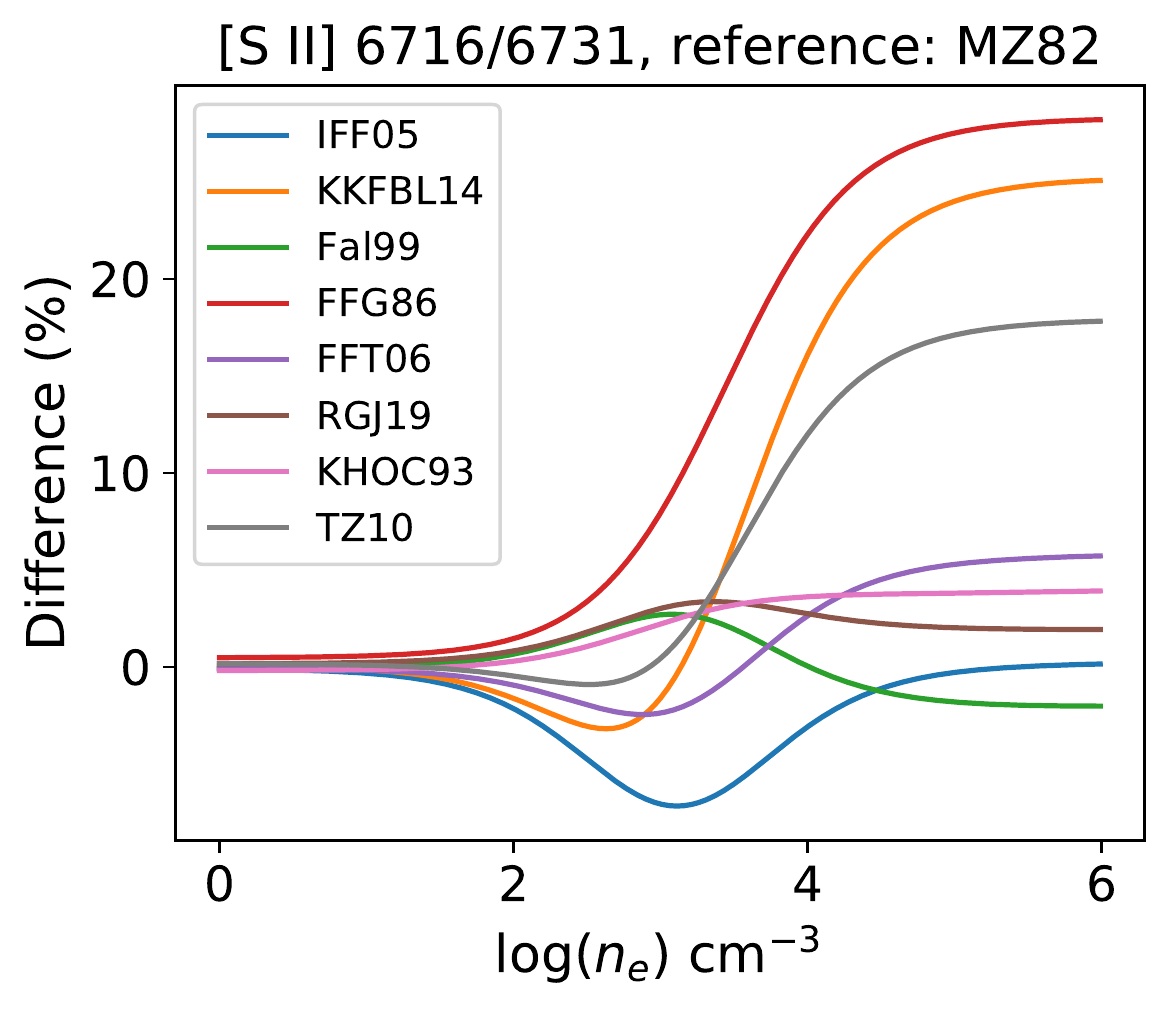}
 \includegraphics[width=0.3\textwidth]{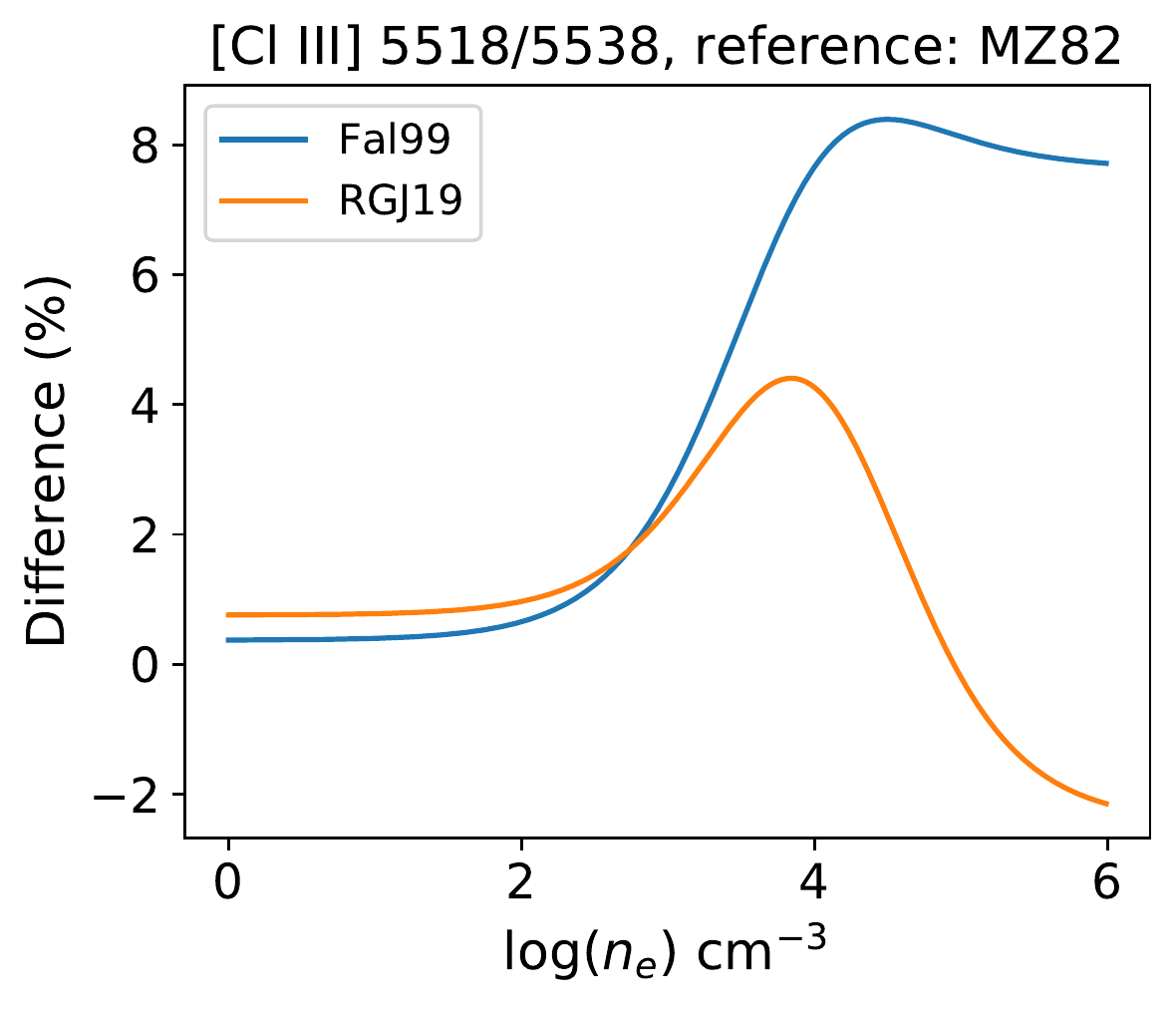}
 \includegraphics[width=0.3\textwidth]{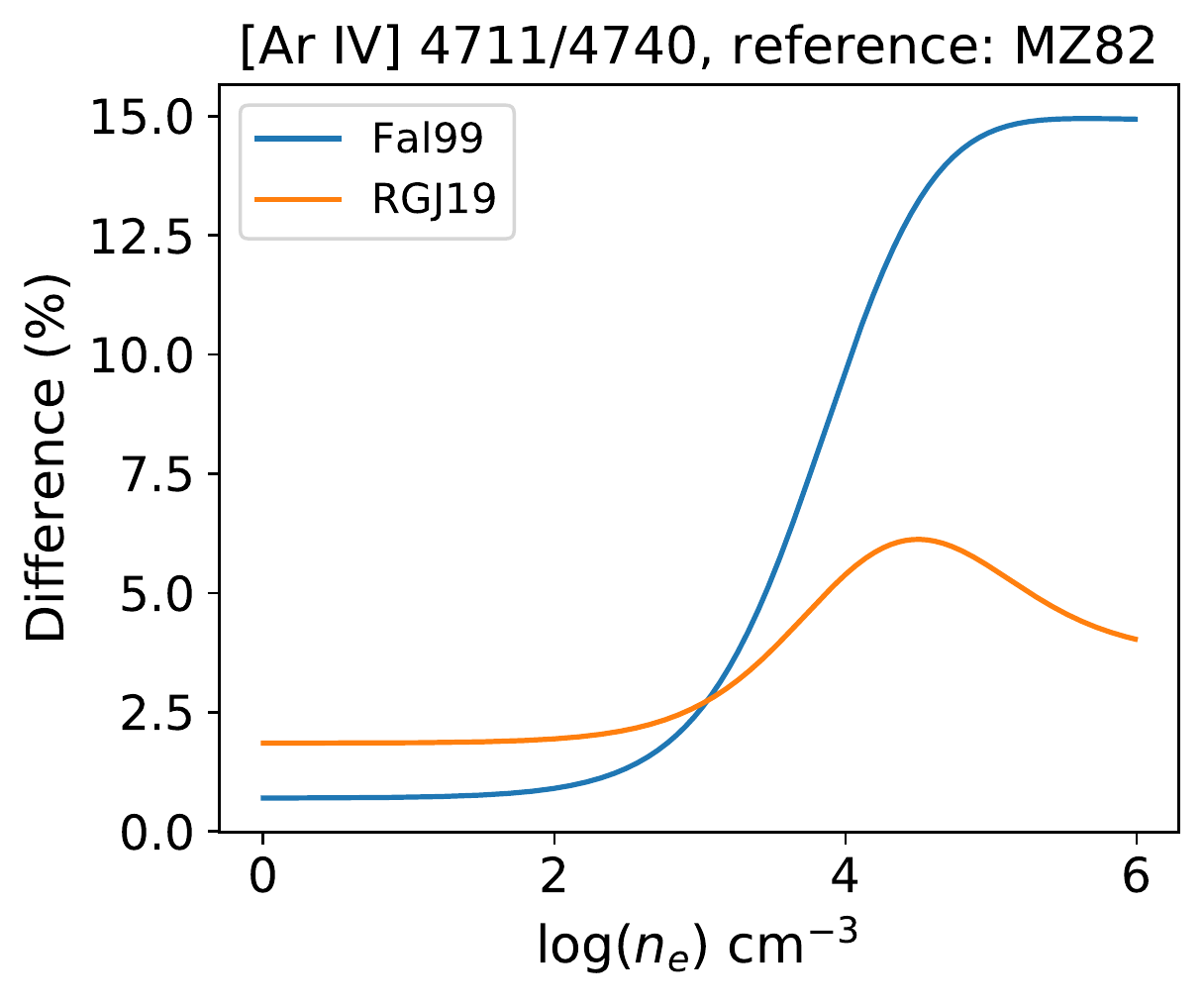}
 \caption{Comparison of the $\varepsilon(^2\mathrm{D^o_{5/2}}-\,^4\mathrm{S^o_{3/2}})/\varepsilon(^2\mathrm{D^o_{3/2}}-\,^4\mathrm{S^o_{3/2}})$ density-sensitive emissivity ratio of P-like ions computed at $T_e=10^4$~K with different $A$-value datasets relative to MZ82. {{Left panel}}: [S~{\sc ii}] $\varepsilon(\lambda 6716)/\varepsilon(\lambda 6731)$. {{Center panel}}: [Cl~{\sc iii}] $\varepsilon(\lambda 5518)/\varepsilon(\lambda 5538)$. {{Right panel}}: [Ar~{\sc iv}] $\varepsilon(\lambda 4711)/\varepsilon(\lambda 4740)$. Source references for the $A$-value datasets are given in Table~\ref{plike}. \label{plike_dif}}
\end{figure}

We repeat this data assessment procedure with the ionic species of the P isoelectronic sequence of nebular interest, namely S~{\sc ii}, Cl~{\sc iii}, and Ar~{\sc iv}. In Figure~\ref{plike_dif}, we plot the percentage difference of the $\varepsilon(^2\mathrm{D^o_{5/2}}-\,^4\mathrm{S^o_{3/2}})/\varepsilon(^2\mathrm{D^o_{3/2}}-\,^4\mathrm{S^o_{3/2}})$ emissivity ratio computed with the FFG86 \cite{fro86}, KHOC93 \cite{kee93}, Fal99 \cite{fri99}, IFF05 \cite{iri05}, FFT06 \cite{fro06}, TZ10 \cite{tay10}, KKFBL14 \cite{kis14}, and RGJ19 \cite{ryn19} $A$-value datasets relative to MZ82 \cite{men82}. Similar to the N-like ions, the overall agreement is around the 25\% level, but if we exclude the three datasets FFG86, TZ10, and KKFBL14 in S~{\sc ii} (see Figure~\ref{plike_dif} {left panel}), the agreement is better than 10\%. The emissivity ratio discrepancies from the fully relativistic (MCDF) Fal99 and RGJ19 datasets relative to MZ82 are puzzling and not explained in \cite{ryn19}: for the former, they increase with atomic number $Z$ up to 15\% in Ar~{\sc iv}, while in the latter they remain within 5\% throughout the isoelectronic sequence (see all {panels} of Figure~\ref{plike_dif}).

In Table~\ref{plike}, we compare the theoretical $R_1$ for the three ions with the line ratios observed in NGC~7027. If we again exclude FFG86, TZ10, and KKFBL14, we find excellent agreement in S~{\sc ii}. Assuming the $A$-value scatter, the predicted electron densities for Cl~{\sc iii} and Ar~{\sc iv} are \mbox{$\log(n_e)=4.76\pm 0.08$}~cm$^{-3}$ and $\log(n_e)=4.73\pm 0.06$~cm$^{-3}$, respectively; i.e., within 0.1~dex. \mbox{From these} comparisons and following our selection criteria, we confidently single out the recent RGJ19 dataset \cite{ryn19} as the {\tt PyNeb} default to ensure a line ratio with an uncertainty from the radiative data better than 10\%. The {\tt PYNEB\_20\_01} dictionary with the {\tt PyNeb} default values adopts \mbox{these datasets.}

\begin{table} [H]
 \caption{Theoretical $A$-values (s$^{-1}$) for the $^2\mathrm{D^o_{3/2,5/2}}-\,^4\mathrm{S^o_{3/2}}$ doublet within the $\mathrm{3s^23p^3}$ ground configuration of P-like S~{\sc ii}, Cl~{\sc iii}, and Ar~{\sc iv}. The $R_1$ ratio is also given for comparison with the line-intensity ratio observed in the dense NGC~7027 planetary nebula. \label{plike}}
 \centering
 \begin{tabular}{cccccc}
 \toprule
 \textbf{Ion} & \textbf{Dataset} & \boldmath{$A(^2\mathrm{D^o_{3/2}}-\,^4\mathrm{S^o_{3/2}})$} & \boldmath{$A(^2\mathrm{D^o_{5/2}}-\,^4\mathrm{S^o_{3/2}})$} & \boldmath{$R_1$} & \textbf{Obs Ratio}\\
 \midrule
 S~{\sc ii} & MZ82 \cite{men82} & 8.82 $\times$ 10$^{-4}$ & 2.60 $\times$ 10$^{-4}$ & 4.42 $\times$ 10$^{-1}$ &  \\
   & FFG86 \cite{fro86} & 6.92 $\times$ 10$^{-4}$ & 2.61 $\times$ 10$^{-4}$ & 5.66 $\times$ 10$^{-1}$ &  \\
   & KHOC93 \cite{kee93} & 8.90 $\times$ 10$^{-4}$ & 2.73 $\times$ 10$^{-4}$ & 4.60 $\times$ 10$^{-1}$ &  \\
   & Fal99 \cite{fri99} & 1.01 $\times$ 10$^{-3}$ & 2.92 $\times$ 10$^{-4}$ & 4.34 $\times$ 10$^{-1}$ &  \\
   & IFF05 \cite{iri05} & 6.84 $\times$ 10$^{-4}$ & 2.02 $\times$ 10$^{-4}$ & 4.43 $\times$ 10$^{-1}$ &  \\
   & FFT06 \cite{fro06} & 7.26 $\times$ 10$^{-4}$ & 2.26 $\times$ 10$^{-4}$ & 4.67 $\times$ 10$^{-1}$ &  \\
   & TZ10 \cite{tay10} & 6.32 $\times$ 10$^{-4}$ & 2.20 $\times$ 10$^{-4}$ & 5.21 $\times$ 10$^{-1}$ &  \\
   & KKFBL14 \cite{kis14} & 5.03 $\times$ 10$^{-4}$ & 1.85 $\times$ 10$^{-4}$ & 5.52 $\times$ 10$^{-1}$ &  \\
   & RGJ19 \cite{ryn19} & 9.43 $\times$ 10$^{-4}$& 2.84 $\times$ 10$^{-4}$ & 4.51 $\times$ 10$^{-1}$ &  \\
   & NGC 7027 \cite{zha05} &  &  &  & 4.43 $\times$ 10$^{-1}$ \\
 Cl~{\sc iii} & MZ82 \cite{men82} & 4.83 $\times$ 10$^{-3}$ & 7.04 $\times$ 10$^{-4}$ & 2.19 $\times$ 10$^{-1}$ &  \\
   & Fal99 \cite{fri99} & 5.04 $\times$ 10$^{-3}$ & 7.91 $\times$ 10$^{-4}$ & 2.35 $\times$ 10$^{-1}$ &  \\
   & RGJ19 \cite{ryn19} & 5.49 $\times$ 10$^{-3}$ & 7.81 $\times$ 10$^{-4}$ & 2.13 $\times$ 10$^{-1}$ &  \\
   & NGC 7027 \cite{zha05} &  &  &  & 2.88 $\times$ 10$^{-1}$ \\
 Ar~{\sc iv } & MZ82 \cite{men82} & 2.23 $\times$ 10$^{-2}$ & 1.77 $\times$ 10$^{-3}$ & 1.19 $\times$ 10$^{-1}$ &  \\
   & Fal99 \cite{fri99} & 2.27 $\times$ 10$^{-2}$ & 2.07 $\times$ 10$^{-3}$ & 1.37 $\times$ 10$^{-1}$ &  \\
   & RGJ19 \cite{ryn19} & 2.34 $\times$ 10$^{-2}$ & 1.93 $\times$ 10$^{-3}$ & 1.23 $\times$ 10$^{-1}$ &  \\
   & NGC 7027 \cite{zha05} &  &  &  & 2.73 $\times$ 10$^{-1}$ \\
 \bottomrule
 \end{tabular}
\end{table}

\subsection{Effective Collision Strengths for C-Like Ions}

\begin{figure}[H]
 \centering
 \includegraphics[width=0.85\textwidth]{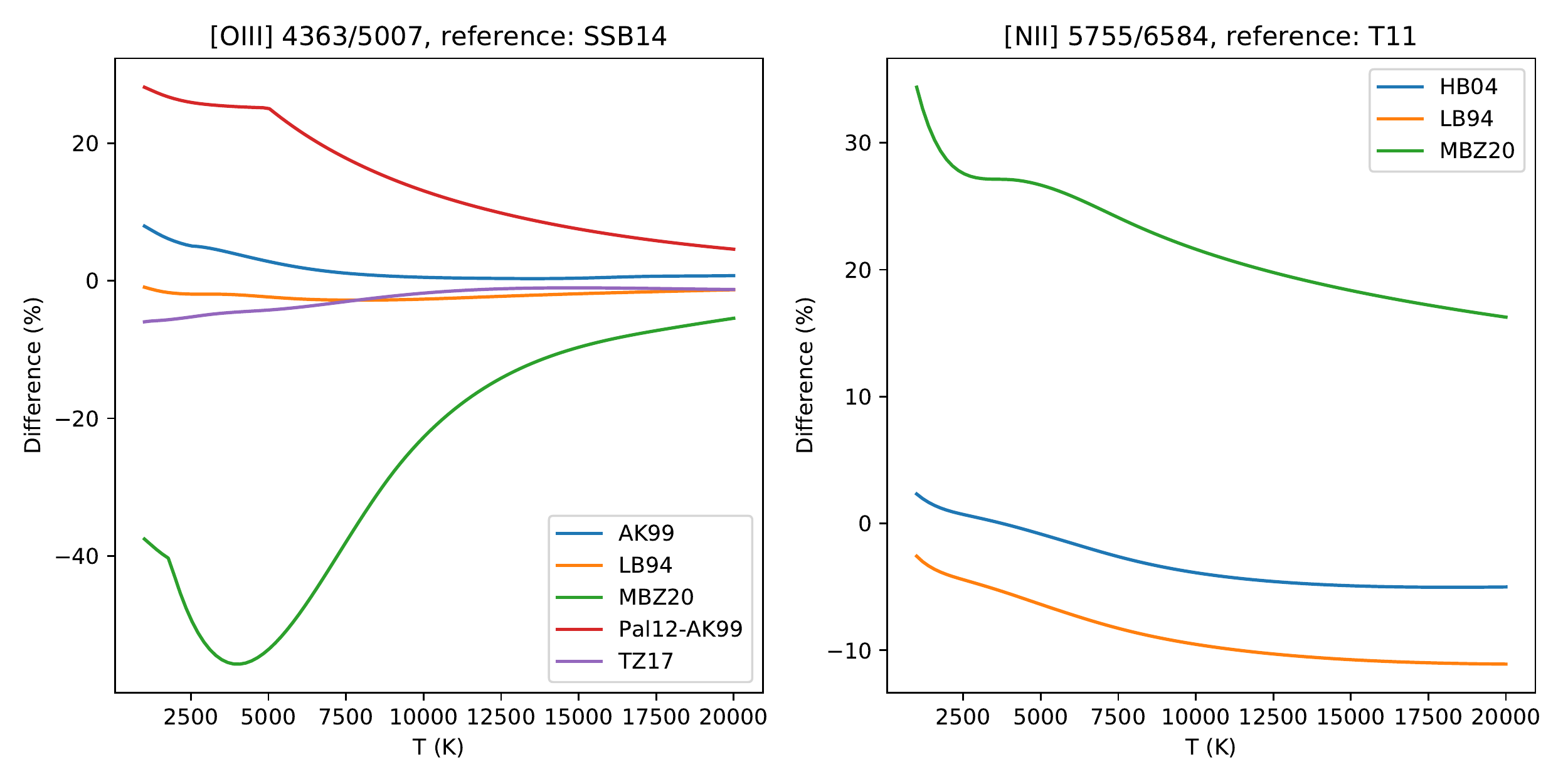}
 \caption{Comparison of the temperature-sensitive emissivity ratios of [O~{\sc iii}] 4363/5007 ({{left panel}}) and [N~{\sc ii}] 5755/6584 ({{right panel}}) computed at $n_e = 10^3$~cm$^{-3}$. For [O~{\sc iii}], we compare the ratios obtained when using the SSB14 default ECS \cite{sto14} with other datasets; for [N~{\sc ii}], the default ECS are T11 \cite{tay11}. Data provenance is: LB94 \cite{len94}; AK99 \cite{agg99}; HB04 \cite{hud04}; Pal12 \cite{pal12}; TZ17 \cite{tay17}; and MBZ20
 \cite{mao20}.} \label{fig:comp_emis}
\end{figure}

The present evaluation of the ECS for ions of the carbon isoelectronic sequence is motivated on the one hand by the leading role of their temperature diagnostics in nebular modeling and on the other by the worrisome discrepancies resulting when using an extensive ECS dataset for C-like ions ($7\leq Z\leq 36$) recently published (MBZ20 \cite{mao20}). In Figure~\ref{fig:comp_emis}, we plot the percentage difference of the [N~{\sc ii}] 5755/6584 and [O~{\sc iii}] 4363/5007 emissivity ratios using the ECS from the LB94 \cite{len94}, HB04 \cite{hud04}, \mbox{and MBZ20} datasets relative to T11 \cite{tay11} (default) for the former and from LB94, AK99~\cite{agg99}, Pal12-AK99~\cite{pal12}, TZ17 \cite{tay17}, and MBZ20 relative to SSB14 \cite{sto14} (default) for the latter. While the accord of LB94, AK99, and TZ17 with SSB14 for O~{\sc iii} and of LB94 and HB04 with T11 for N~{\sc ii} is well within
10\% for both species, MBZ20 shows large discrepancies: as high as $-$50\% in O~{\sc iii} and +30\% in N~{\sc ii}. As

\noindent
also shown in Figure~\ref{fig:comp_emis} ({left panel}), the emissivity ratio resulting from the Pal12-AK99 ECS for O~{\sc iii} is $\sim$25\% higher and is caused by their neglecting the 2p$^4kl$ free channels in the close-coupling expansion of the ion--electron system that leads to an energy downshift of the broad 2p$^5$ resonance \cite{sto14}.

\begin{figure}[H]
 \centering
 \includegraphics[width=0.9\textwidth]{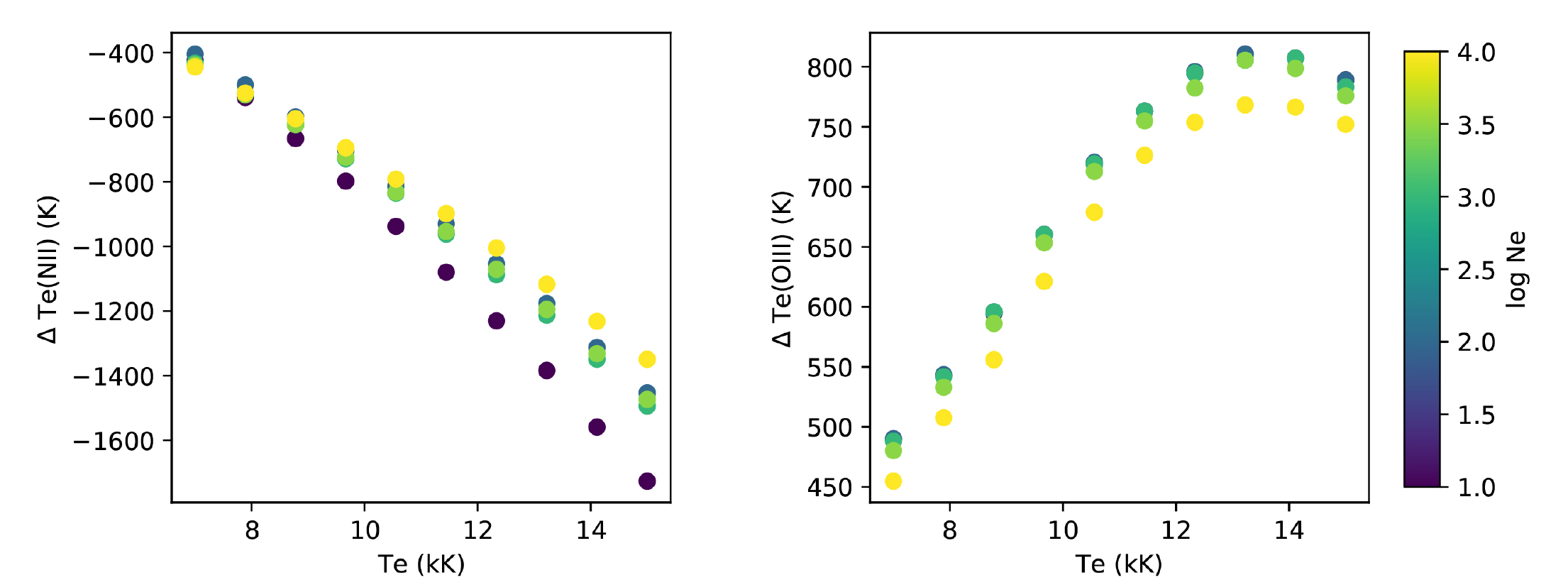}
 \caption{Temperature differences caused by using the MBZ20 ECS relative to the {\tt PyNeb} defaults. {{\mbox{Left panel}}}: $T_e$(N~{\sc ii}). {{Right panel}}: $T_e$(O~{\sc iii}).}
 \label{fig:temps_Mao}
\end{figure}

The large discrepancies in the emissivity ratios originating from the MBZ20 ECS are of concern and demand further analysis. For instance, in Figure~\ref{fig:temps_Mao}, we plot the temperature differences when using MBZ20 relative to the default ECS datasets: \new{for a set of temperatures and densities, we compute the diagnostic line ratios [N~{\sc ii}] 5755/6584 and [O~{\sc iii}] 4363/5007, and from them, we compute back the electron temperature determined using MBZ20.} $T_e$(N~{\sc ii}) is underestimated by as much as 1.6~kK at 14~kK, while $T_e$(O~{\sc iii}) is overestimated by 0.8~kK. This temperature divergence may have a significant impact on the ionic abundances since $T_e$(N~{\sc ii}) is used in nebular models for the singly ionized species (N$^{+}$, O$^{+}$, and S$^{+}$), while $T_e$(O~{\sc iii}) is used for O$^{++}$, Ar$^{++}$, and S$^{++}$.

\begin{figure}[H]
 \centering
 \includegraphics[width=0.9\textwidth]{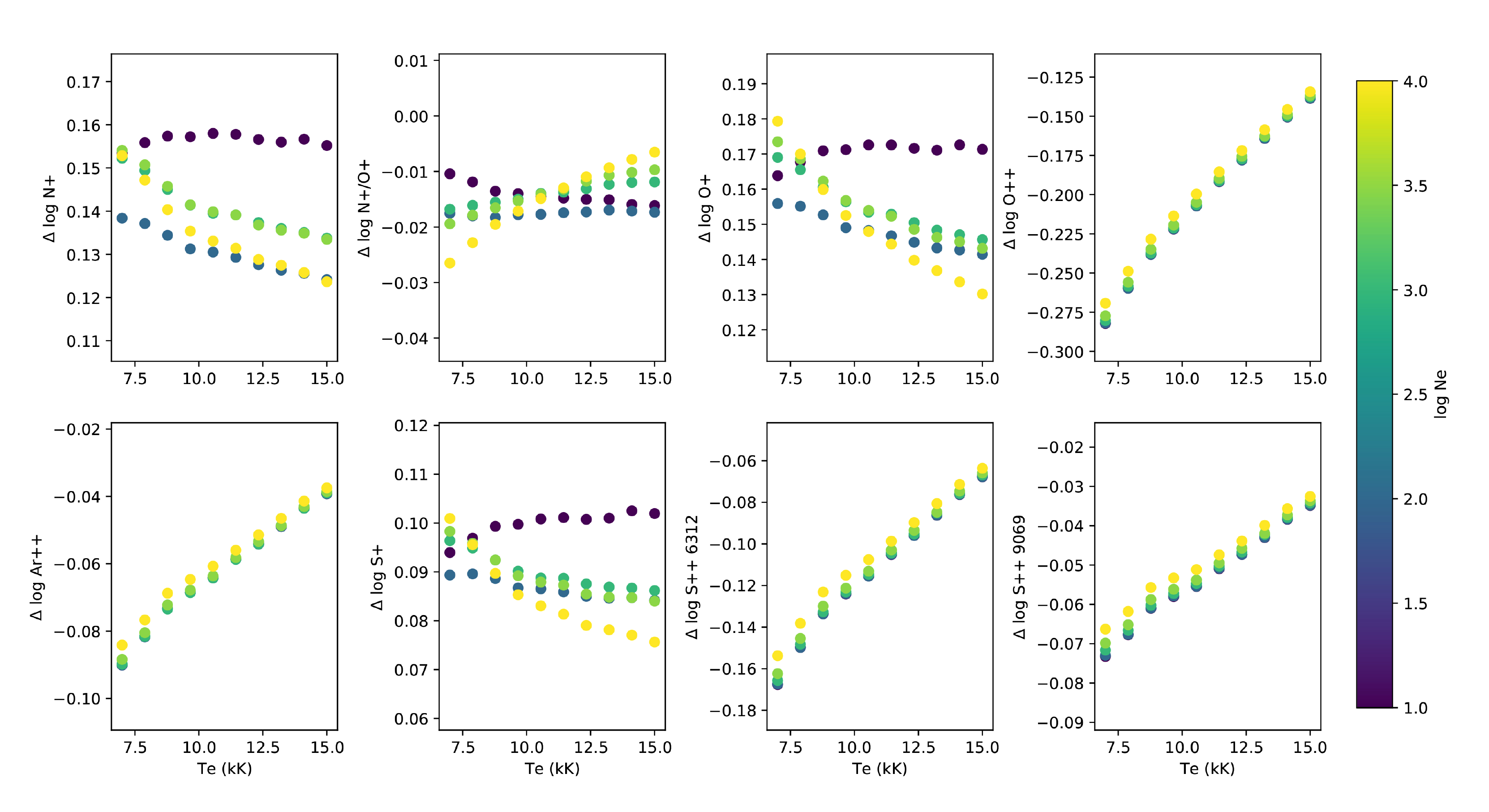}
 \caption{Abundance logarithmic differences for several ions caused by using the MBZ20 ECS relative to the {\tt PyNeb} defaults.}
 \label{fig:ionic_Mao}
\end{figure}

In Figure~\ref{fig:ionic_Mao}, we show the resulting abundance logarithmic differences: \new{we compute the emissivities of the emission lines with the {\tt PyNeb} default atomic data and then use them to compute back the ionic abundances with MBZ20}. They are obtained at five electron densities (10, 100, 1000, 3000, \mbox{and 10,000 cm$^{-3}$}) from the following emission lines: [N~{\sc ii}] 6584; [O~{\sc ii}] 3726 + 29; [O~{\sc iii}] 5007; [Ar~{\sc iii}] 7135; \mbox{[S~{\sc ii}] 6716 + 31;} and [S~{\sc iii}] 6312 and 9069. The differences (positive and negative) can be as large as 0.3 dex when considering low or high ionization species. Depending on the global ionization of the nebula, \mbox{the total} abundance may be underestimated (high-ionization nebulae) or overestimated (low-ionization nebulae) depending on the dominant ionic stage of each element. \mbox{For medium-ionization} nebulae, \mbox{the net result} may be a cancellation between both effects, leading to the same elemental abundances.

\begin{figure}[H]
 \centering
 \includegraphics[width=1.0\textwidth]{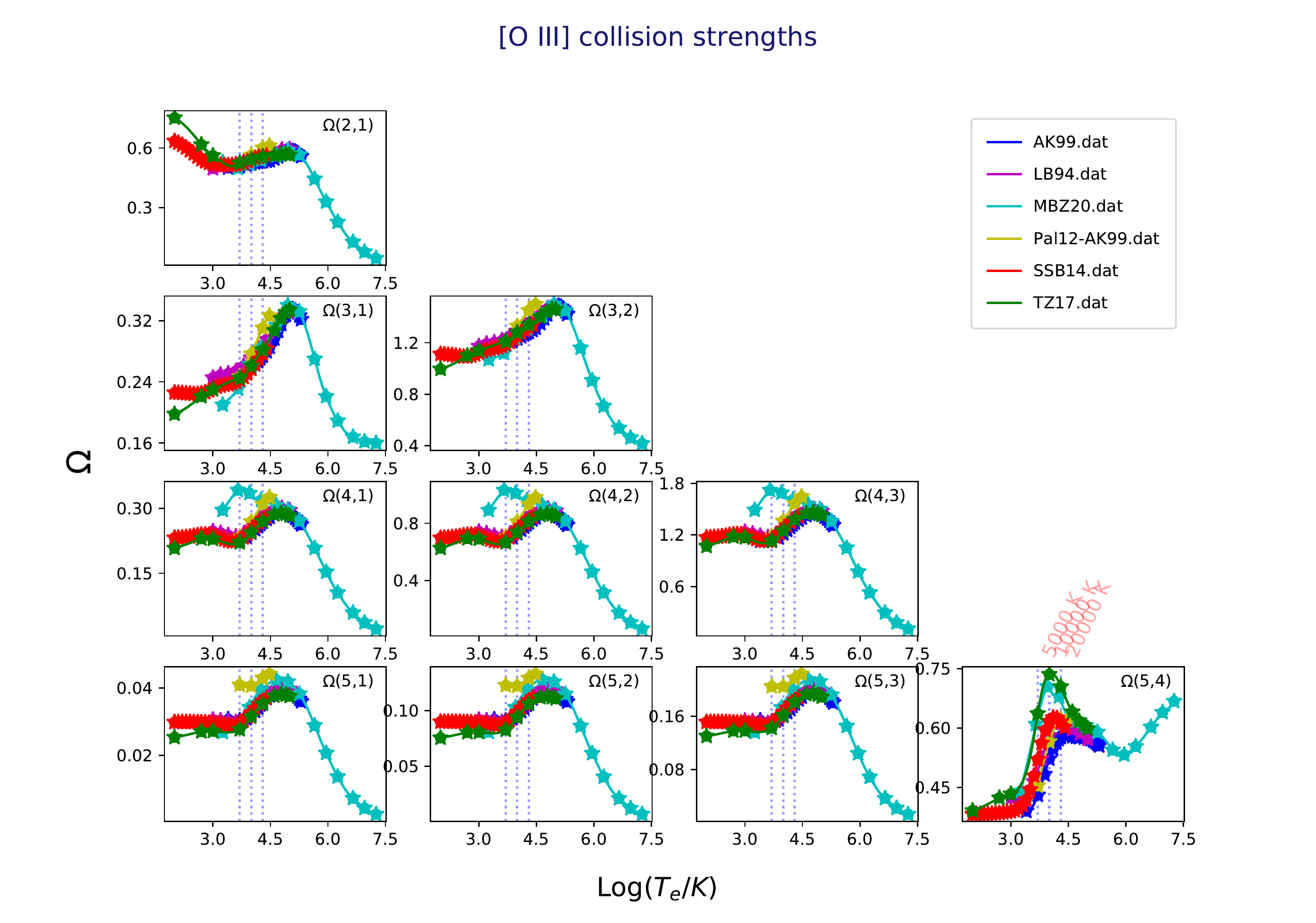}
 \caption{Available ECS for O~{\sc iii} in {\tt PyNeb} plotted with the {\tt plotOmega} method of the {\tt DataPLot} class. \newnew{Note that ``collision strengths'' and $\Omega(k,i)$ in this figure stand for temperature-dependent ECS}.}
 \label{fig:o3_cs}
\end{figure}

\begin{figure}[H]
 \centering
 \includegraphics[width=1.0\textwidth]{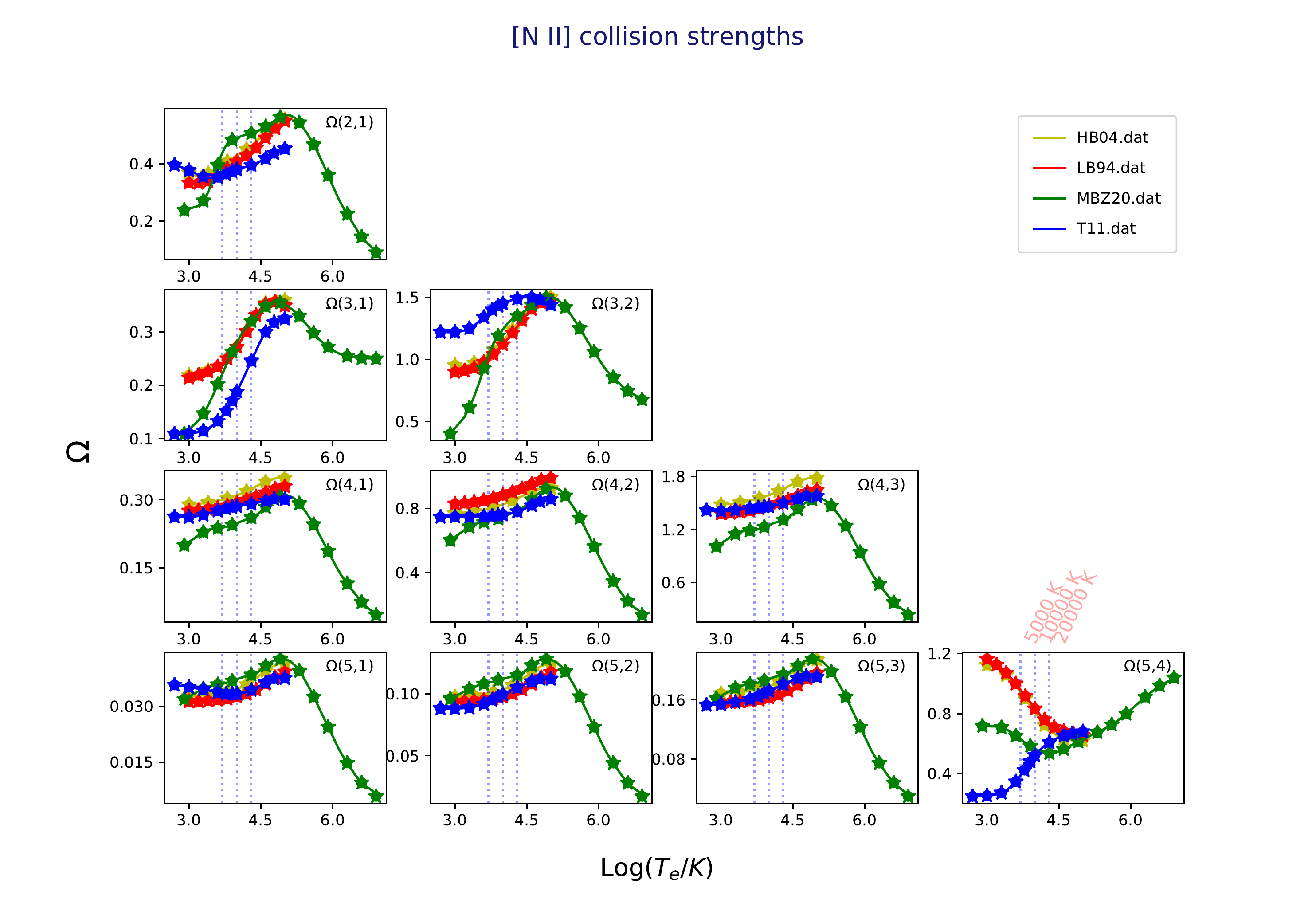}
 \caption{Available ECS for N~{\sc ii} in {\tt PyNeb} plotted with the {\tt plotOmega} method of the {\tt DataPLot} class. \newnew{Note that ``collision strengths'' and $\Omega(k,i)$ in this figure stand for temperature-dependent ECS}.}
 \label{fig:n2_cs}
\end{figure}

The {\tt DataPlot} class offers a series of useful methods to plot the atomic data, which can be used to revise the MBZ20 datasets for O~{\sc iii} and N~{\sc ii}:
\begin{verbatim}
 dp_O3=pn.DataPlot('O',3,NLevels = 5)
 dp_O3.plotOmega(figsize = (10,8))
 dp_N2=pn.DataPlot('N',2,NLevels = 5)
 dp_N2.plotOmega(figsize = (10,8))
\end{verbatim}
the resulting plots being shown in Figures~\ref{fig:o3_cs} and \ref{fig:n2_cs}. For O~{\sc iii}, it may be seen that the $\Omega(4,1)$, $\Omega(4,2)$, and $\Omega(4,3)$ ECS by MBZ20 corresponding to the transitions $\mathrm{2s^22p^2\ ^1D_2-\,^3P_J}$ are considerably larger in the temperature region of interest ($5\leq T_e\leq20$~kK). On the other hand, the situation in N~{\sc ii} for these transitions is the inverse: the MBZ20 ECS are now lower. The $\Omega(5,4)$ from different datasets for both O~{\sc iii} and N~{\sc ii} also displays incongruent behaviors. However, we underline that MBZ20 is the only dataset offering ECS for temperatures $\log(T_e)\gtrsim 5.5$~K.

Figures~\ref{fig:o3_cs} and \ref{fig:n2_cs} illustrate a recurrent problem in nebular modeling when dealing with ECS tabulations. The temperature end-points and mesh intervals in each dataset vary, and they must be somehow interpolated. \new{In {\sc chianti}, the ECS are sometimes represented by five, six, or seven points, two of which are assigned to $T_e=0$ and $T_e=\infty$; thus, in real terms, the finite temperature range is reduced to three, four, or five point spline fits, which may result in further inaccuracies.}

\begin{figure}[H]
 \centering
 \includegraphics[width=0.4\textwidth]{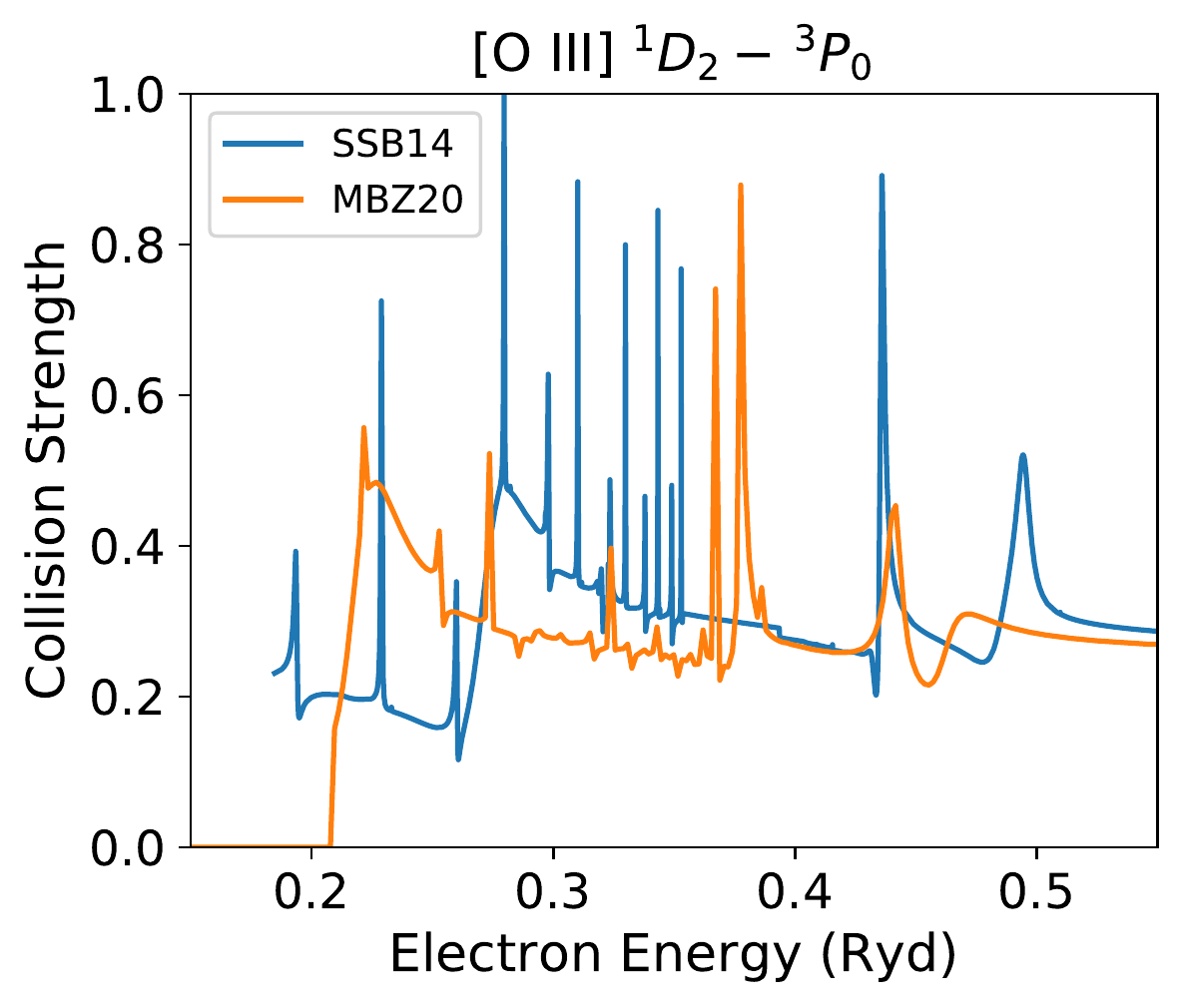}
 \includegraphics[width=0.4\textwidth]{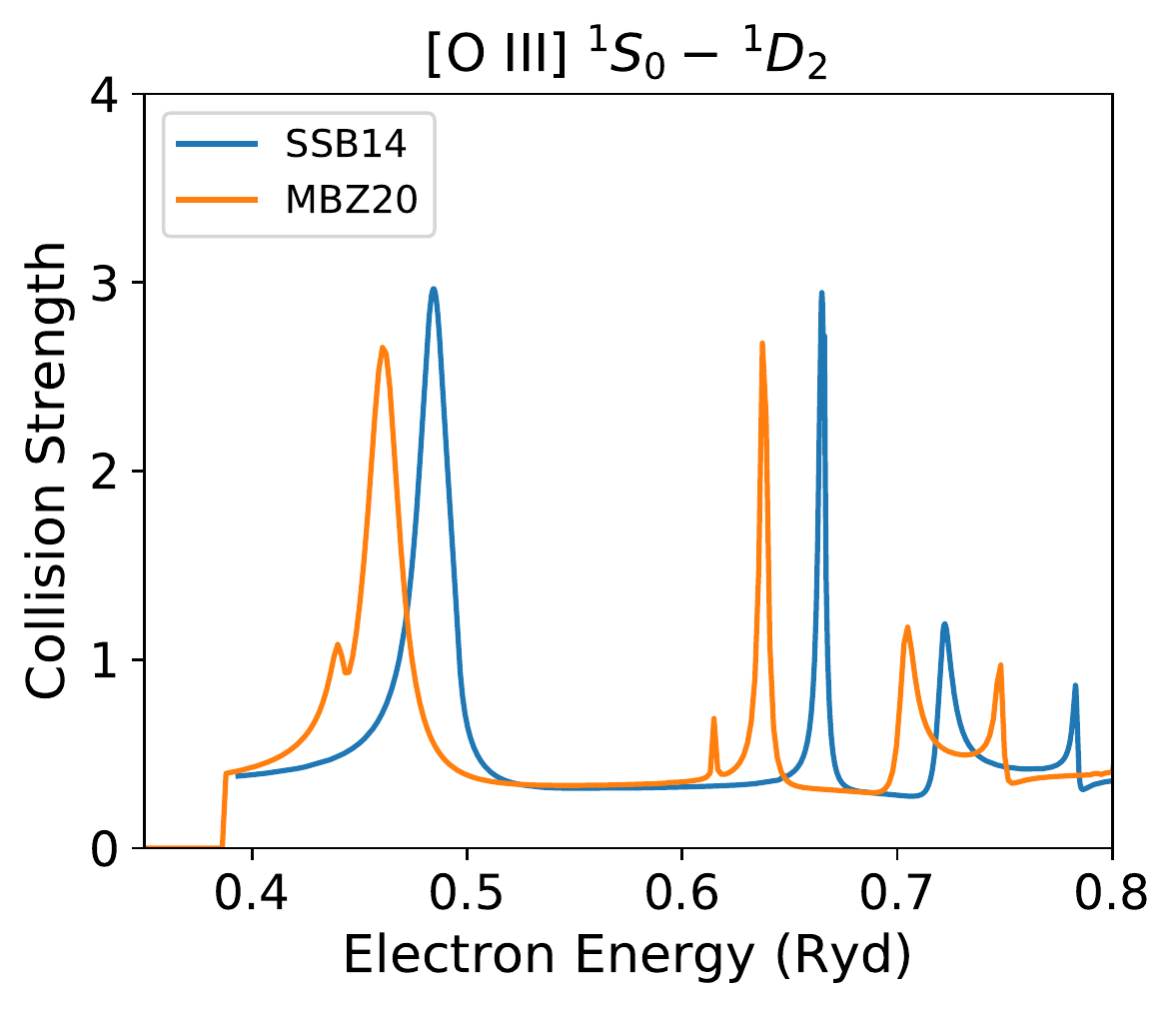}
 \caption{Collision strength for the transitions [O~{\sc iii}] $\mathrm{^1D_2-\,^3P_0}$ ({{left panel}}) and $\mathrm{^1S_0-\,^1D_2}$ ({{right panel}}) computed by SSB14 and MBZ20.}
 \label{o3_om}
\end{figure}

In Figure~\ref{o3_om} ({left panel}), we plot the collision strength for the [O~{\sc iii}] $\mathrm{^1D_2-\,^3P_0}$ transition computed by MBZ20 and SSB14, which is dominated in the near-threshold region by a broad peak. In SSB14, this peak is located at $\approx$0.29~Ryd, while in MBZ20, it downshifts to $\approx$0.22~Ryd, causing the ECS enhancement shown in Figure~\ref{fig:o3_cs}. As discussed by \cite{tay17}, there is an experimental position for this peak at 0.294~Ryd \cite{nii02}; \new{thus, the downshifted resonance position in MBZ20 is inaccurate. This peak is mainly caused by an array of resonances with configuration 2s2p$^3$3s, and their atomic structure does not represent these states adequately}. Moreover, the resonance structure in the two curves in the energy interval $0.35\leq E\leq 0.55$~Ryd is also different (see Figure~\ref{o3_om}, left panel).

The $\Omega(5,4)$ ECS in Figure~\ref{fig:o3_cs} corresponds to the [O~{\sc iii}] $\mathrm{^1S_0-\,^1D_2}$ transition, and the data by SSB14 and MBZ20 are also found to be discrepant in the temperature region of interest. \mbox{As shown} in Figure~\ref{o3_om} ({right panel}), this discord is again caused by resonance positions, in this case the near-threshold 2p$^5$ resonance that in MBZ20 is around 0.02~Ryd lower. Although this downshift is relatively smaller, the resonance is closer to the threshold, making the ECS sensitive at low temperatures. Moreover, it is also seen in Figure~\ref{fig:o3_cs} that the $\Omega(5,4)$ by MBZ20 and TZ17 are in agreement. We therefore predict that the position of this resonance in TZ17 is also lower than in SSB14, and we would then be of the opinion that the latter dataset is the most accurate and, therefore, the {\tt PyNeb} default. Similar discrepancies are observed in the $\Omega(5,4)$ in N~{\sc ii} (see Figure~\ref{fig:n2_cs}), which were discussed in \cite{men14} and need further evaluation.

\section{Discussion and Conclusions}

In the context of nebular modeling, we made use of the atomic data assessment capabilities of {\tt PyNeb} to revise the accuracy of the $A$-values for transitions within the ground configuration of the N- and P-like ions as suggested by \cite{jua17} and the ECS for C-like species motivated by the recent publication of an extensive dataset \cite{mao20}. For atomic data evaluation, {\tt PyNeb} offers two coveted features: a fairly complete historical database of radiative and collisional parameters and a handy series of class methods to compare plasma diagnostics and abundances using the different atomic datasets, \mbox{which enable} direct uncertainty estimates in parameter space. As a result, we have been able to select for these isoelectronic sequences the reference datasets to become the {\tt PyNeb} defaults.

The magnitude and scatter of the lifetimes of the two S~{\sc ii} 3s$^2$3p$^3\ ^2\mathrm{D^o_{3/2,5/2}}$ levels have been extensively reviewed recently to propose a viable measurement scheme to improve their accuracy to better than 10\% \cite{tra20}. It was mentioned therein that a 10\% lifetime uncertainty would lead to \mbox{a 50\%} emissivity ratio uncertainty and that the current theoretical scatter is larger than 10\%. Due to inherent experimental difficulties to measure the very long lifetimes of the $^2\mathrm{D^o_{3/2,5/2}}$ levels ($\sim$1~h), \mbox{the more} feasible task of measuring the lifetimes of the $^2\mathrm{P^o_{1/2,3/2}}$ levels (a few seconds) was developed. \mbox{In the} present work, we showed that theory can do better than 50\%: for both the N- and P-like ions of nebular interest, the emissivity ratio uncertainties despite the $A$-value scatter are within 25\%. Furthermore, by means of the high-density benchmark with the observed line ratios (i.e., from the NGC~7027 planetary nebula), we can discard $A$-value datasets to reduce the emissivity ratio uncertainty to the 10\% level with some degree of assurance. However, we notice that, in spite of accurate $R_1$ ratios, the absolute $A$-values in some datasets are somewhat larger, particularly those for [O~{\sc ii}] $A(^2\mathrm{D^o_{5/2}}-\,^4\mathrm{S^o_{3/2}})$ recently computed with the MCDF method \cite{che07, han14, he18}. This type of discrepancy can only be resolved by measurement, although we remain skeptical in light of the long-standing discord between theory and experiment regarding the Fe~{\sc xvii} 3C/3D $f$-value ratio \cite{kuh20}.

Regarding the recent MBZ20 ECS for the carbon sequence \cite{mao20}, we demonstrated that this dataset leads to large discrepancies in the emissivity ratios of both N~{\sc ii} and O~{\sc iii} and should therefore not be used in the electron temperature range $5\leq T_e\leq 20$~kK. As shown in Figure~\ref{o3_om}, the broad peak at $\approx$0.29~Ryd in the O~{\sc iii} $\Omega(\mathrm{^1D_2,^3P_0})$ collision strength by SSB14 \cite{sto14} (default) is downshifted in MBZ20 to $\approx$0.22~Ryd, in disagreement with an experimental benchmark of 0.294~Ryd \cite{nii02,tay17}. \mbox{The origin} of this discrepancy may be found in the large O~{\sc iii} atomic model (590 fine-structure levels from \mbox{24 configurations} including orbitals with principal quantum number $n\leq 5$) of MBZ20, which has been optimized specifically to provide ECS at high temperatures. \new{They adjusted the $nl$-dependent scaling parameters in a systematic way without manual re-adjustment to avoid introducing arbitrary changes across the isoelectronic sequence. This strategy leads to poor atomic structure for low-charge ions such as O~{\sc iii} and, thus, less accurate collision data at low temperatures. At high temperatures, \mbox{the resonance} effect is reduced, so their data should be more accurate} (Mao Junjie, private communication). As a result, we can conclude that the uncertainty level in the [N~{\sc ii}] 5755/6584 emissivity ratios using ECS from LB94, HB04, and T11 and in [O~{\sc iii}] 4363/5007 from LB94, AK99, SSB14, and TZ17 is within 10\%. This outcome is a good example that the latest and largest atomic calculation is not necessarily the most accurate and should therefore be carefully examined, as we showed here, before data are included in plasma models. This last point has also been stressed in the context of the long-standing discrepancies in the ECS computed for Be-like ions by independent $R$-matrix methods \cite{agg17}.

\new{After providing and applying atomic data in nebular modeling for almost five decades, \mbox{we are} amazed at not being able to guarantee an acceptable level of accuracy. However, the present work contributes to clarifying that computations for low- and high-energy regimes cannot be treated simultaneously with the same level of accuracy, a fact that must be taken into account in atomic database maintenance. We tend to agree with \cite{tra20} that we have a laboratory astrophysics problem in hand with no foreseeable solution due to the endemic difficulties in computing, measuring, \mbox{and evaluating} the data products. As proposed in \cite{men14b}, open and fluid user--provider interactions and early data curation schemes in the research cycle are perhaps the best we can do.}

As future work within the present initiative, we intend to revise and complete the atomic data for the lowly ionized species of the iron-group elements, for which there are non-conclusive accuracy evaluations. We would finally like to mention that, in reference to data-curation strategies, a historical atomic database as kept in {\tt PyNeb} is a welcome step in data preservation, and the choice of a flat ASCII format rather than FITS to facilitate longevity and user interaction must be seriously considered in database design.

\vspace{6pt}



\authorcontributions{C.M. (Christophe Morisset) and C.M. (Claudio Mendoza) conceptualized and coordinated the full paper. Investigation, methodology, analysis, and validation, J.G.-R., M.A.B., and V.G.-L.; software, V.L., J.G.-R., and C.M. (Christophe Morisset); writing, the original draft preparation, V.L., J.G.-R., \mbox{C.M. (Christophe Morisset),} and C.M. (Claudio Mendoza); writing, review and editing, J.G.-R., M.A.B., C.M. (Christophe Morisset), and \mbox{C.M. (Claudio Mendoza);} visualization, V.L., V.G.-L., C.M. (Christophe Morisset), and C.M. (Claudio Mendoza); ancillary calculations, M.A.B. All authors read and agreed to the published version of the manuscript.}

\funding{C.M. (Christophe Morisset) and V.G.-L. acknowledge support from the Mexican CONACyT-CB2015-254132 and UNAM-DGAPA-PAPIIT-101220 projects. C.M. (Claudio Mendoza) is grateful for financial support from the NASA Astrophysics Research and Analysis Program (Grants 12-APRA12-0070 and 80NSSC17K0345). J.G.-R. acknowledges support from an Advanced Fellowship from the Severo Ochoa excellence program (SEV-2015-0548) and from the State Research Agency (AEI) of the Spanish Ministry of Science, \mbox{Innovation and} Universities (MCIU) and the European Regional Development Fund (FEDER) under Grant AYA2017-83383-P. J.G.-R. also acknowledges support under Grant P/308614 financed by funds transferred from the Spanish Ministry of Science, Innovation and Universities, charged to the General State Budgets and with funds transferred from the General Budgets of the Autonomous Community of the Canary Islands by the MCIU.}

\acknowledgments{We are grateful to Dr. Mao Junjie (Strathclyde University, U.K.) for giving us access to the raw collision strengths of \cite{mao20} and for helpful and clarifying discussions \new{and to Dr. Elmar Tr\"abert (Ruhr-Universit\"at Bochum, Germany) for enlightening comments on measuring the immeasurable. We are indebted to Dr. Gra\.{z}yna Stasi\'nska (Observatoire de Paris, France) for regular suggestions to improve {\tt PyNeb} in the last decade, \mbox{especially during} the preparation of the NEBULATOM workshops. We thank these three colleagues for critically reading the manuscript and providing useful comments and addenda that led to its improvement.}}

\conflictsofinterest{The authors declare no conflict of interest.}

\abbreviations{The following abbreviations are used in this manuscript:\\

\noindent
\begin{tabular}{@{}ll}
 FITS & Flexible Image Transport System \\
 NIST & National Institute of Standards and Technology \\
 QED & Quantum electrodynamics
\end{tabular}
}
\reftitle{References}





\end{document}